# Quelques Aspects de la Dynamique
# des Milieux Granulaires

## P. Evesque
Lab MSSMat, UMR 8579 CNRS, Ecole Centrale Paris
**92295 CHATENAY-MALABRY, France,** e-mail: evesque@mssmat.ecp.fr

**Résumé:**

*On décrit la diversité du comportement dynamique des matériaux granulaires à l'aide de quelques exemples simples. On aborde le problème du gaz granulaire et on montre qu'il n'existe qu'à l'état très peu dense ;à plus forte concentration il devient inhomogène (clustering). Est-ce une transition de phase? On aborde le problème du billard de Sinaï vibré. On étudie la propagation des ondes ultrasonores, et leur diffusion (simple et multiple) par les grains lorsque la longueur d'onde est petite. On montre que des sollicitations quasi statiques cycliques peuvent engendrer un écoulement plastique et on relie ce problème (i) à celui du comportement rhéologique (contrainte-déformation) quasi statique et (ii) à celui des écoulements engendrés par vibration. On montre que l'on peut engendrer de la houle à l'interface liquide sable, grâce à des vibrations horizontales très intenses. Enfin on aborde le problème de la ségrégation. On rappelle, en particulier, qu'un mouvement tridimensionnel complexe (brassage) est nécessaire pour bien mélanger .*

**Pacs # : 5.40 ; 45.70 ; 62.20 ; 83.70.Fn**

La dynamique des milieux granulaires présente beaucoup de facettes distinctes; et c'est un domaine de recherche très actif actuellement dans les différents domaines scientifiques: physique, mécanique, géophysique, chimie, pharmacie,… Le nombre de travaux publiés est tel que l'on ne peut les citer tous, encore moins les décrire, ou les critiquer, ce que devrait faire tout travail scientifique. Le but poursuivi ici ne sera donc pas celui-là. Je décrirai essentiellement deux ou trois problèmes distincts, qui me semblent illustrer la position actuelle du débat scientifique, tout en reflétant ses avancées et ses limites et pour lequel un transfert de technologie me semble possible, permettant ainsi soit le développement de nouveaux diagnostiques, soit l'amélioration des techniques expérimentales. Je considérerai un milieu granulaire soumis à des sollicitations cycliques variables et montrerai qu'il peut se comporter de façons très différentes. Nous verrons que le comportement d'une bille dans une boîte vibrée n'est pas ergodique; c'est en partie du à la dissipation. Puis on considérera l'exemple d'un milieu granulaire dans l'espace et on montrera sa tendance naturelle à se densifier; c'est probablement d'un intérêt crucial pour la cosmologie, mais cela montre aussi que la forme naturelle du milieu granulaire est un empilement relativement dense, c'est-à-dire formant un réseau continu.

Le premier "vrai" problème que nous aborderons ensuite est celui de la propagation des ondes sonores et ultrasonores dans un milieu granulaire. Les ondes que nous considérerons seront de faibles amplitudes pour qu'elles ne perturbent pas le réseau de contacts en se propageant. Ces ondes lorsqu'elles se propagent traversent un matériau qui est hétérogène à l'échelle mésoscopique puisqu'il est formé de grains différents, de taille micro- ou milli-métrique. Se posent alors différents problèmes qui touchent à ceux (i) de la localisation des ondes (localisation d'Anderson), (ii) de la diffusion simple, (iii) de la diffusion multiple, (iv) de la cohérence et des interférences des ondes diffusées. Une expérience récente nous permettra d'illustrer ces questions.

Le deuxième problème est celui de la génération d'un écoulement sous des conditions d'excitation périodique: On verra qu'on peut engendrer l'écoulement d'un milieu granulaire, en lui imposant des cycles de contrainte-déformation, même lorsque ceux-ci sont périodiques, de petites amplitudes et correspondent à des conditions quasi statiques. Pour ce faire, il faut que les déformations qu'on engendre périodiquement soit de nature plastique; il faut donc qu'elles soient plus grandes que dans le cas précédent, concernant la propagation ou de la diffusion des ondes. On fera le parallèle entre ce problème et celui de l'"acoustic streaming" que les ingénieurs rencontrent lorsqu'ils soumettent un liquide contenu dans un récipient à des ondes sonores de forte intensité, ou qui pousse les poussières au nœud ou au ventre de vibration dans un tube de Kundt [0.1]. Ceci montrera la différence qualitative de comportement qu'engendre le remplacement du frottement visqueux par du frottement "sec" (solide). Mais ceci nous permettra aussi d'aborder les problèmes de chaos déterministe et de cinétique de mélange, qui sont des problèmes pratiques importants tant pour les milieux granulaires que plus généralement pour tout milieu fluide. C'est aussi un problème capital en génie chimique ou en génie des procédés. Nous prolongerons ce thème par un exemple d'étude de ségrégation granulaire dans la quatrième partie.

Avant ce dernier exemple, nous continuerons à décrire le comportement fluide des milieux granulaires, mais nous soumettrons ce milieu à des sollicitations encore plus intenses et rapides. Dans ce cas, les forces inertielles seront tellement fortes qu'elles domineront le comportement et la mécanique; on pourra ainsi négliger les lois rhéologiques. Nous montrerons alors qu'un milieu granulaire se comporte comme un liquide parfait (*i.e.* sans viscosité et sans frottement). Ce type de comportement, même s'il est rare sur Terre, peut devenir prépondérant en apesanteur, car un milieu granulaire, même compacte, peut n'être soumis à aucune contrainte latérale dans ce cas.

Comme nous l'avons déjà signalé, le dernier problème abordé est celui de la ségrégation car il pose le problème de l'écoulement des milieux granulaires, de la définition des forces ou des mécanismes de mélanges (diffusion, convection,…); nous reprendrons alors le concept de l'advection chaotique développé dans l'exemple 2 pour montrer qu'un bon mélangeur doit générer des écoulements tri-dimensionnels. Nous étudierons ce phénomène de ségrégation dans le cas particulier du turbula; c'est un mélangeur tridimensionnel faisant subir au milieu granulaire des mouvements périodiques tridimensionnels complexes et engendrant un écoulement près de la surface libre. Nous montrerons que les forces de mélange produites par l'advection chaotique ne sont pas suffisantes pour lutter contre les forces de ségrégation; par contre, nous verrons aussi qu'on peut réduire l'efficacité de la ségrégation en augmentant les forces inertielles et en les rendant multi-directionnelles, car celles-ci permettent de "modifier en permanence la direction du champ gravitaire" et de l'aligner sur la direction de l'écoulement.

## 1. Dynamique de la matière granulaire en apesanteur:

Sous quelle forme peut-on trouver la matière granulaire en apesanteur? Une réponse au moins partielle à cette question peut être donnée maintenant puisque nous avons eu la chance de mener une expérience en micro-gravité en février 1998 à l'aide d'une fusée sonde Minitexus 5, partie d'Esrange (Norvège) [1.1-1.2]. Cette expérience, financée essentiellement par l'ESA, et par un support CNES pour les missions, consistait à imposer une vibration sinusoïdale polarisée linéairement d'amplitude a et de fréquence $\omega/2\pi$ variables (a=0.3mm-2.5mm , $\omega/2\pi$=3Hz-60Hz) à trois cellules cubiques d'1cm$^3$ contenant des billes de bronze en quantité respective, moins d'une couche de billes, 2 couches de billes et 3 couches de billes.

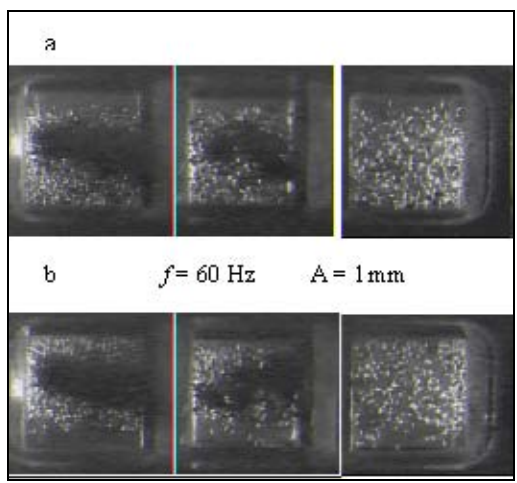

*Figure 1.1: Billes vibrées en apesanteur* ; *fréquence 60Hz ,amplitude 1mm. De gauche à droite, cellule contenant : 3 couches de billes, 2 couches, 1 couche. La direction de vibration est « verticale ».*
*Photos a: cellule en position la plus "élevée".*
*Photos b: cellule en position la plus "basse".*
*On observe (i) la formation d'un amas dense entouré d'un gaz lâche dans le cas des deux cellules les plus denses, (ii) l'existence d'un gaz lâche dans la cellule la moins dense, (iii) le décollement du gaz de la paroi du haut (figures du haut) ou du bas (figures du bas), ce qui montre que l'excitation est « supersonique ».*

Comme le montre la Figure 1.1, l'observation du comportement des billes dans les cellules a démontré qu'un milieu granulaire vibré en apesanteur ne peut avoir un comportement «gazeux» que lorsque la densité de ses particules est suffisamment faible pour que le libre parcours moyen $l_c$ de chaque grain soit limité par la taille L du container vibrant et non par la présence des autres billes. Dans ces conditions le gaz est extrêmement raréfié et les grains sont presque isolés dans la boîte. Ce régime correspond pour un gaz réel au régime dit de *Knudsen* ($l_c$>L*)*. Dès que la probabilité moyenne de rencontre entre grains augmente et dépasse un, on observe la formation d'un amas dense qui reste plus ou moins immobile dans le container; cet amas est pris en sandwich entre deux couches de "gaz de Knudsen", très peu dense; des mécanismes d'"évaporation" et de "condensation" des grains par l'amas assurent l'équilibre des différentes phases.

De plus, on a observé que le container se détache du gaz périodiquement au cours du mouvement (ceci est visible sur la Fig. 1.1); ceci indique que la vitesse moyenne des grains de la phase gazeuse est inférieure à la vitesse maximum du container; en d'autres termes, les conditions d'excitation sont *supersoniques* [1.2] et le container joue le rôle d'un vélostat, et non celui d'un thermostat.

Enfin l'étude de la statistique des impacts des billes sur un capteur de pression situé sur la face "haute" de la cellule la moins dense a permis de mesurer une pression moyenne P qui varie avec les conditions d'excitation. On a montré que P variait comme $(a\omega)^{1.5}$. L'application de la statistique de Boltzmann prédit que cette pression doit être proportionnelle au carré de la vitesse moyenne des billes v². Nous pensons que la différence de variation entre v et aω est un artefact lié au temps τ de réponse du capteur par rapport au nombre de collisions par seconde $N_s$ qu'il reçoit: lorsque $N_s\tau >1$, la pression $P\approx N_s\tau v$ mesurée est $P\approx N_s\tau v$, et est donc proportionnelle à v²; tandis que lorsque $N_s\tau <1$ la pression mesurée est soit 0, soit proportionelle à v.

Ces résultats sont encore trop partiels pour qu'ils puissent être considérés comme scientifiquement prouvés et nous poursuivons ce type d'expérience. Mais ils semblent cohérents. De plus, ils ont été partiellement confirmés par des approches numériques et théoriques [1.3-1.4]; en particulier, l'effondrement du gaz en un amas dense est un domaine de recherche théorique actif.

Enfin, ces résultats démontrent le caractère très particulier du gaz granulaire, et ce au moins à plusieurs titres:

i) ce gaz ne peut exister que lorsque les grains sont quasi isolés (régime de Knudsen); ceci est lié au caractère dissipatif des collisions granulaires. Dans ces conditions, il est difficile de parler de la thermodynamique de l'état granulaire gazeux, car la densité ρ d'un tel gaz doit tendre vers 0 lorsque le volume est grand: si d est le diamètre des grains, la densité ρ ne peut excéder ρ<1/(Ld²). On voit ainsi que certaines variables thermodynamiques qui devraient être extensives ne le sont pas. En fait, il vaudrait mieux parler de la "thermodynamique du grain isolé" ou de la "particule isolée"; en d'autres termes, ce problème de "gaz granulaire" est très proche de celui du billard proposé par Sinaï [1.5]. Rappelons que ce dernier est un des piliers de la théorie du chaos, et peut donner lieu à l'élaboration d'une physique statistique particulière. Nous en donnons une description succincte à la fin de cette section. (En fait nos expériences récentes nous ont démontré que la trajectoire d'une bille vibrée pouvait ne pas être ergodique, à cause de la dissipation).

ii) Il est intéressant de rappeler que la description mécanique d'un gaz (d'atomes) à partir des équations de la mécanique des milieux continus ne s'applique que pour des volumes plus grand que le libre parcours moyen $l_c$. (Par exemple, les équations de propagation du son ne s'appliquent pas dans le cas d'un gaz de Knudsen). Le fait qu'un gaz granulaire n'existe qu'en régime de Knudsen prouve donc qu'il ne peut obéir aux équations de la mécanique des milieux continus. Ceci laisse présager l'existence de grandes difficultés pour décrire la mécanique des écoulements granulaires denses.

iii) cette expérience montre que l'excitation du milieu granulaire est en général du type "supersonique"; c'est probablement lié au caractère dissipatif des collisions. Mais ceci n'est pas sans conséquence, car l'on sait qu'une excitation supersonique donne lieu à des équations différentielles de type hyperbolique, qui admettent des discontinuités; dans ces conditions il est peut-être difficile de supposer la continuité des variables d'état en tout point du milieu. Ceci est probablement l'une des raisons majeures de la difficulté de prédire le comportement des fluides granulaires dans les régimes où les collisions dominent les processus de transfert et de dissipation.

iv) L'utilisation de la notion de température granulaire $T_g$ est peut-être dangereuse. En effet, cette expérience montre que le paramètre important qui contrôle la vitesse v des particules est la vitesse aω de la boîte. La boîte doit donc être considérée comme un *vélostat* et non comme un *thermostat* [1.6]: 2 systèmes de particules ayant le même nombre de couches auront la même vitesse v mais pas la même « température granulaire » $T_g=mv^2/2$. D'où des difficultés d'équilibre thermique local en cas de mélange, ce qui peut expliquer le phénomène de ségrégation.

v) Le fait que le milieu granulaire gazeux ne peut pas être décrit par des équations de la mécanique des milieux continus est difficile à mettre en évidence sur terre, car les grains sont naturellement confinés par le fond du container, ce qui fait que la densité du milieu varie continûment avec la hauteur dans l'échantillon. On ne voit donc aucune discontinuité apparaître. Ce phénomène peut quand-même être mis en évidence sur terre, comme nous l'avons montré en étudiant le mécanisme d'instabilité du « démon de Maxwell » dans les gaz granulaires [1.17]; l'expérience confirme la valeur $n_c\approx 1$ du nombre critique de couche.

Toujours est-il que cette expérience démontre l'efficacité du processus de "mise en tas" ; ainsi le « tas » est pratiquement l'"état naturel" d'un milieu granulaire. Par la suite nous ne considérerons plus que des situations où le matériau granulaire est dense, les grains en contact les uns avec les autres. Mais pour l'instant diluons ce gaz.

### *Billard granulaire :*

Comme nous l'avons dit, le régime de Knudsen est le régime des particules sans interaction ; il doit donc être très proche du cas limite d'une bille dans une boîte. Nous avons donc simulé numériquement ce cas limite lorsque la boîte est à 1 dimension [1.8] ; puis nous avons fait l'expérience 3d en apesanteur (campagnes Airbus de Mars & Septembre 2002) ; dans ce cas, le container était cylindrique et immobile, et la bille agitée par un piston vibrant (a sin[ωt]). L'expérience a permis de montrer la pertinence du modèle 1d car il y a stabilisation de la trajectoire de la bille sur un axe perpendiculaire à la direction de vibration, (la position de cet axe semble aléatoire, mais varie très lentement) ; on observe aussi la mise en synchronisation du mouvement de la bille sur le mouvement du piston lorsque l'amplitude a du mouvement du piston est grande par rapport à la taille de la boîte L (1/2>a/L>0.01) ce qui a été souvent le cas expérimentalement. Le mouvement de la bille est alors périodique à la fréquence $\nu=\omega/(2\pi)$.

Pour comprendre la raison de la pertinence du modèle 1d, des simulations numériques 3d par la méthode des éléments discrets ont été réalisées au laboratoire à l'aide du code de J.J. Moreau et M. Jean ; le modèle tient compte des frottements solides et des divers coefficients de restitution ; le container est cubique et la direction de vibration est perpendiculaire à une face. Les simulations confirment les résultats expérimentaux. Elles montrent de plus que le mouvement converge très vite vers un mouvement linéaire, dont la trajectoire est stable et s'oriente le plus souvent parallèlement au mouvement de la boîte : il y a donc brisure spontanée de l'ergodicité de la trajectoire, ce qui explique la pertinence du modèle unidimensionnel.

L'existence de frottement aux parois permet aussi de stabiliser des trajectoires perpendiculaires à la direction de vibration, lorsque les conditions initiales le permettent ; l'obtention d'un tel mouvement montre l'existence de bassins d'attraction différents suivant les conditions initiales.

Une trajectoire stable, parallèle au mouvement de vibration, n'est pas obligatoirement parfaitement périodique ; elle peut apparaître de façon intermittente (chaotique) aux faibles amplitudes de vibration; elle ne devient périodique qu'à plus forte amplitude. On constate aussi expérimentalement que les trajectoires géométriques restent stables autour d'une position moyenne si l'on choisit bien la forme du piston : cette trajectoire reste stable, même lorsqu'on la déstabilise en désorientant (de 10° à peu près) la direction de vibration par rapport à l'axe de la cellule, ou par rapport à la normale aux facettes du cube. Ces trajectoires stables sont perpendiculaires aux faces et non parallèles à la direction de vibration, ou à la direction de la gravité.

Cette stabilité explique la *brisure d'ergodicité*. On peut analyser le phénomène à partir de la *théorie du billard* : On sait qu'un billard cubique a des trajectoires fermées ; en ce sens il n'est pas ergodique ; ces trajectoires peuvent être vues comme des « modes propres » ; la vibration alimente en énergie certaines de ces trajectoires et entretient le mouvement lorsque la dissipation est suffisamment faible. Dans le cas contraire le mode s'éteint et n'est pas observé. Les trajectoires perpendiculaires aux faces et parallèles à la direction de vibration sont donc des modes simples fermés, facilement entretenus. La dimension de l'espace des phases (qui permet de décrire le mouvement) se réduit et passe de 11-d (2*2 rotation+2*3 translation + temps) à 1-d.

La forme de la cavité joue sûrement un rôle important sur l'existence des « modes propres » et sur l'ergodicité du problème. Il est probable qu'une cavité en forme de calotte sphérique fermée par un fond plat ne devrait pas présenter une telle séparation en « modes propres ». De même l'adjonction d'obstacles fixes à l'intérieur de l'enceinte devrait améliorer la qualité de l'ergodicité ». Ainsi, peut-être que l'existence seule des autres billes rend le problème du gaz granulaire relativement ergodique. Mais ceci reste réellement à démontrer.

## 2. Propagation d'ondes acoustiques et ultrasonores dans un milieu granulaire:

On sait que lorsqu'on soumet un milieu granulaire à des ondes sonores ou ultrasonores de suffisamment basse fréquence, ces ondes se propagent dans le milieu. La propagation de ces ondes est un phénomène bien connu de la vie courante: elle explique pourquoi on ressent le passage du métro ou d'un train dans les maisons situées aux abords des gares. Les indiens d'Amérique s'en servent pour déduire la distance et le nombre de cavaliers arrivant à leur rencontre…

D'un point de vue plus scientifique, l'existence de ces ondes est liée à l'existence des modules élastiques $K$ et de cisaillement $\mu$. On sait déterminer ces modules au laboratoire sur des échantillons de sable de petite taille, par exemple par la méthode de la colonne résonante, ou en mesurant la réponse à de toute petite déformation. Sur le terrain, on peut utiliser des stimuli vibratoires ou des explosions pour déterminer les vitesses des ondes de compression et de cisaillement à l'aide de capteurs piézo-électriques, ou des sismographes; on utilise alors la relation entre ces vitesses du son et les modules élastiques pour déterminer ces derniers:

$$V_p = [(K + 4/3\mu)/\rho]^{1/2} \quad \text{et} \quad V_s = (\mu/\rho)^{1/2} \tag{2.1}$$

Où ρ est la densité du matériau. On peut ainsi reconstituer les différentes strates géologiques, leur nature, leur épaisseur, leur orientation,… Et ces mesures concordent entre elles.

Plus récemment [2.1], des mesures à l'échelle locale ont été réalisées par Liu et Nagel sur un ensemble de grains soumis au seul champ de pesanteur. Une impulsion sonore de fréquence 4kHz était émise par un haut parleur; l'onde sonore recueillie quelques (*i.e.* 10-15) grains plus loin par un microphone de la forme et de la taille d'un grain (0.7cm) était ensuite analysée. L'amplitude A du signal reçu était constituée d'un montée rapide puis d'une traînée qui fluctuait rapidement et décroissait ensuite lentement ; ce signal était constitué d'une partie haute fréquence (> 4kHz) et d'une partie basse fréquence (<4kHz). Cette expérience conclut qu'il y avait une ambiguïté dans la mesure de la vitesse du son. En effet le temps de vol (ou d'arrivée du front d'onde) permettait de définir une vitesse du son beaucoup plus rapide que la vitesse de groupe $\partial A/\partial \omega$ mesurée par l'analyse du signal complet. De plus l'expérience montrait une sensibilité extrême à des variations même très faibles de la configuration des billes; enfin le spectre de Fourier du signal recueilli présentait une loi de puissance.

Cette expérience a posé quelques problèmes d'interprétation à la communauté des physiciens qui voyait là l'archétype d'un nouveau type de comportement propre aux matériaux granulaires. Ils ont alors fait le lien avec une expérience plus ancienne, due à Dantu [2.2], reprise par De Josselin de Jong, puis par Travers et al. [2.2]; cette expérience montra que les contraintes se propageait de façon inhomogène à l'échelle de quelques grains et qu'il existait un réseau de chaînes de forces sous-jacent. Une partie de la communauté physicienne a ainsi été amenée à penser que ce réseau de forces pouvait être inhomogène à toute échelle de longueur. Elle a alors interprété les résultats de propagation du son de Liu et Nagel comme étant liés à la caractéristique inhomogène du réseau des forces de contact et à la nature "fragile" [2.3] de la mécanique des milieux granulaires, nature fragile liée d'après eux à l'existence de ce réseau inhomogène de forces et de contacts.

Cette nouvelle doctrine est-elle défendable? L'expérience ci-dessus [2.1] remet-elle en cause l'approche classique des mécaniciens? Doit-on la reconsidérer en profondeur? C'est ce que nous voulons discuter dans cette section. Dans la prochaine sous-section nous donnerons de nouveaux éléments expérimentaux [2.4] qui réconcilieront au moins partiellement les deux points de vue; ils montreront que lorsque l'onde est de fréquence très basse, elle se propage comme une onde acoustique classique; cependant, lorsque sa fréquence augmente, et que sa longueur d'onde diminue, sa propagation peut engendrer des phénomènes de diffusion simple ou multiple donnant lieu à des figures d'interférence (speckle) [2.5], qui ralentissent la traversée du milieu, qui disperse l'onde aussi dans le temps et qui atténue l'amplitude de l'onde directe. Mais les caractéristiques de l'onde diffusée peuvent permettre de définir de nouvelles techniques de mesure et d'analyse des mouvements des grains.

En fait, le problème de la propagation des ondes dans un milieu hétérogène peut devenir plus complexe que ne laissent voir les expériences que nous allons décrire;

elle peut donner lieu en particulier au mécanisme de "localisation des ondes", appelées localisation d'Anderson [2.6 , 2.7]. Ces phénomènes sont importants, généraux et complexes; leur champ d'application est vaste et dépasse la mécanique des matériaux; on peut penser en effet aux propagations des ondes sismiques, aux ras-de-marées,… [2.7].

### 2.1. *Expériences récentes de propagation d'onde ultrasonore dans un matériau granulaire:*

Quelques années après l'étude de Liu et Nagel [2.1], une expérience de propagation du son plus propre et plus facile à interpréter a été conçue [2.4] dans un oedomètre; elle utilisait une émission à plus haute fréquence (1MHz), des capteurs piézo-électriques et un nombre de grains plus grand. Elle a permis d'élucider les phénomènes observés par Liu et Nagel [2.1] et de les rattacher aux phénomènes de propagation du son et de diffusion multiples.

La Fig. 2.1 représente le montage expérimental et des résultats typiques lorsque le signal émis par l'émetteur est une impulsion de 2μs centrée sur 500kHz et de bande passante 20kHz-1MHz. Le signal reçu par le détecteur est constitué dans l'ordre d'arrivée, d'une onde directe (notée E), d'une onde réfléchie (R), puis d'un train d'onde (S) très long correspondant au signal générée par les diffusions multiples de l'onde principale. A partir du retard entre l'émission et l'arrivée du signal (E), on peut calculer la vitesse du son V. Les variations de V en fonction de P sont données dans la Fig. 2.2; une valeur typique est V=750m/s à P=0.75 MPa, ce qui correspond à une longueur d'onde $\lambda \approx 1.5$mm. La variation de V avec P est approximativement normale, reflétant la forme sphérique des grains et la présence de contacts de type Hertz, *cf.* Légende de la Fig.2.2 .

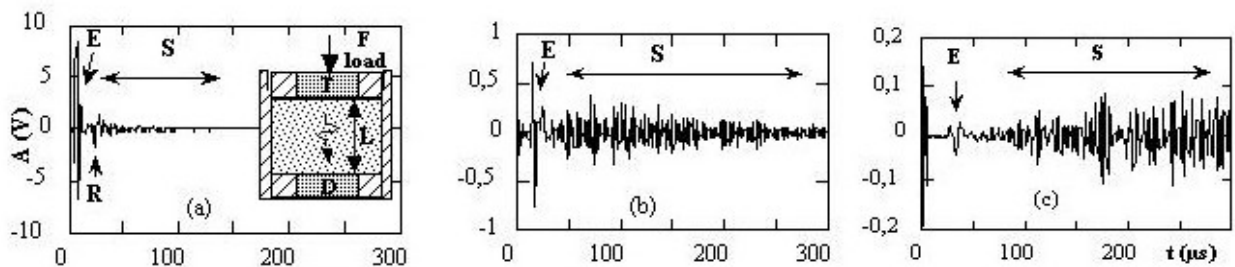

*Figure 2.1:* Signal ultrasonore mesuré par un transducteur piézo-électrique de 12mm *de diamètre dans un empilement de billes de verre polydisperses (diamètre* d) *sous un chargement oedométrique à* P=0.75 MPa . *(a)* d = 0.2 - 0.3 mm, *(b)* d = 0.4 - 0.8 mm, *(c)* d = 1.5 ± 0.15mm.
(E) *et* (R) *correspondent respectivement à la propagation simple (cohérente) du signal sonore émis par la source et à sa réflexion sur les murs du haut et du bas.* (S) *est un signal dû aux diffusions multiples. Le schéma expérimental est donné dans l'insert de (a):* T *et* D *sont les émetteur et récepteur.* [2.4]

L'analyse du spectre de Fourier de l'onde transmise (E) pour les différents empilements montre l'existence d'une fréquence de coupure haute fréquence $\omega_c/2\pi$ qui dépend de la taille des grains et de la vitesse de propagation de l'onde E; on trouve expérimentalement la relation $\omega_c d \approx 2\pi V$; $\omega_c$ correspond donc à une longueur d'onde $\lambda=$ $2\pi V/\omega_c$ égale au diamètre d du grain. Les fréquences $\omega/2\pi$ plus grandes que $\omega_c/2\pi$ ne sont pas transmises, mais diffusées.

La Fig. 2.1 montre que le signal diffusé est d'autant plus faible que la taille du grain est plus petite, à longueur d'onde donnée; ceci est un comportement normal comme nous le rappellerons par la suite.

De même on constate sur la Fig. 2.1 que plus le diamètre D du récepteur est petit, plus le signal diffusé est important. Ceci est un phénomène comparable à ce qui se passe en optique du speckle: le speckle est une figure d'interférence lumineuse produite par l'interférence des ondes diffusées en un point particulier, issues de la même source (nombre d'onde $k_o$), mais diffusées (nombre d'onde **k**) par différents diffuseurs i en position $\mathbf{r_i}$ et arrivant à l'instant t au point **r** considéré; cette interférence produit une intensité qui varie dans l'espace sur une échelle de longueur $l_s$ typique appelée longueur de corrélation. En effet l'intensité globale locale liée à ces interférences varie suivant que les interférences de ces ondes sont constructives ou destructives au point considéré et dépend donc des déphasages $\mathbf{k_i(r-r_i)}+\mathbf{k_o r_i}$ entre les ondes $k_i$ issus des différents diffuseurs localisés en $r_i$ et interférant en r. L'intensité locale

$$A= \{\Sigma_i A_i \exp[-i\{\mathbf{k_i(r-r_i)}+\mathbf{k_o r_i}\}] \exp[i\omega t]\} \quad (2.2)$$

est donc une variable aléatoire de longueur de corrélation typique $l_s$. Son intégration sur la surface du capteur donne l'amplitude du signal reçu par le détecteur; cette intégrale est aussi une variable aléatoire; mais ses fluctuations relatives sont réduites par le fait que c'est une moyenne de variables aléatoires.

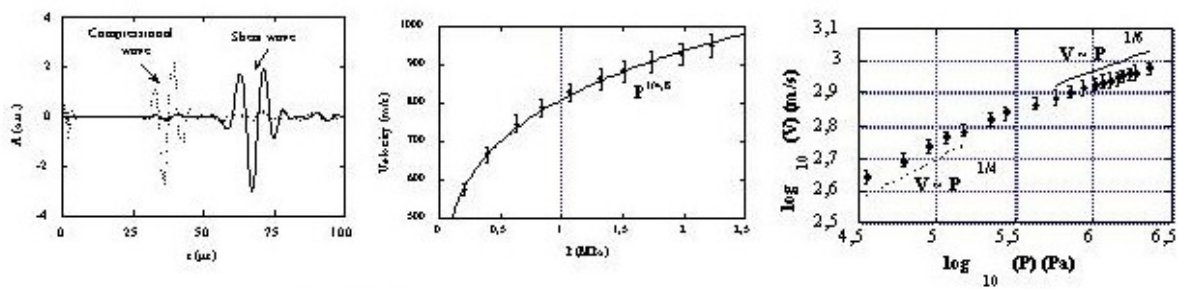

***Figure 2.2:*** *A partir du retard* $\delta T$ *entre le signal reçu E et le signal source de la figure 2.1, on peut déduire la vitesse de propagation V des ondes sonores E de compression (ou de cisaillement qui ne sont pas représentées ici) et ses variations en fonction de la contrainte* P, *ici pour les billes* d = 0.4 - 0.8 mm. *On constate que V suit la loi prévue par la théorie de Hertz lorsque la contrainte est suffisante, mais qu'elle en dévie à basse contrainte* P. *On peut attribuer cette déviation à l'accroissement du nombre réel de contacts effectifs quand on augmente* P *à faible* P. *[2.4]*

Revenons aux signaux de la Fig. 2.1; nous avons vu que le signal diffusé ne devient important que pour la deuxième et troisième expérience (Figs. 2.1.b et 2.1.c); dans la Fig.2.1.a le signal E est dominant, la diffusion faible, ce qui est en accord avec une estimation simple de la section efficace de diffusion $\sigma_D$ de l'onde Rayleigh qui varie comme $\sigma_D \sim (d/\lambda)^4$. Cependant, le fait que l'amplitude diffusée devienne faible quand la cellule reste petite (taille L), n'implique pas pour autant que la diffusion ne joue pas un rôle essentiel dans la propagation des ondes à longue distance, c'est-à-dire dans des échantillons beaucoup plus grands. En effet, comme nous l'avons dit, on peut rendre compte du phénomène de diffusion en définissant la section efficace de diffusion $\sigma_D$ du matériau pour l'onde considérée; un autre paramètre important, qui est lié à $\sigma_D$ et à la densité de diffuseurs $\rho_D$, est le libre parcours moyen $l_c$ entre deux processus de diffusion. La relation qui lie ces trois grandeurs est: $\rho_D\, l_c\, \sigma_D = 1$. Quelque soit le matériau, le phénomène de diffusion deviendra le phénomène prépondérant dès que $l_c < L$, où L est la taille du matériau. En particulier, l'intensité $I_E$ de l'onde directe (E) qui traverse l'épaisseur r diminue avec cette épaisseur r suivant l'équation probabiliste:

$$I_E = I_{E_o} \exp(-r/l_c) \qquad (2.3)$$

Cette équation ressemble à une équation d'absorption; en fait, ce n'est pas le cas; elle indique que les diffuseurs rendent opaque le milieu. C'est un phénomène bien connu, qui a son analogue en optique: elle explique par exemple pourquoi un nuage plus épais laisse moins passer la lumière qu'un autre plus mince. Cette équation indique donc que la lumière n'arrive pas sur la Terre; elle n'indique pas que la lumière est absorbée par le nuage; en fait, la lumière est rétro-diffusée vers l'extérieur de la Terre. Lorsque l'émetteur est placé à l'intérieur du milieu, et que ce milieu est très diffusant ($l_c << L$), toute l'onde émise doit sortir soit par l'avant, soit par l'arrière du matériau. Cependant, si l'excitation dure un temps très bref, l'émission du matériau peut durer beaucoup plus longtemps car le trajet réel de l'onde est très variable et dépend à l'instant t du nombre de processus élémentaires de diffusion réalisés.

Si le milieu absorbe aussi l'onde, il faut adjoindre un second terme à l'Eq. (2.3): $I_E = I_{E_o} \exp(-r/l_c - \alpha r)$, où $\alpha$ est le coefficient d'absorption. Une manière d'estimer les deux termes est de mesurer l'intensité rétro-diffusée, provoquée par une excitation à l'extérieure de l'échantillon; en effet, lorsque le milieu est très diffuseur, l'intensité rétro-diffusée, intégée sur toutes les directions du demi-espace, doit être égale à l'intensité émettrice; dans le cas contraire, elle sera atténuée d'un facteur $\exp(-2\alpha l_c)$ puisque l'onde rétro-diffusée aura parcourue en moyenne une longueur $2l_c$ dans le milieu. Bien entendu, il faut tenir compte de l'intensité sonore directement réfléchie par la surface.

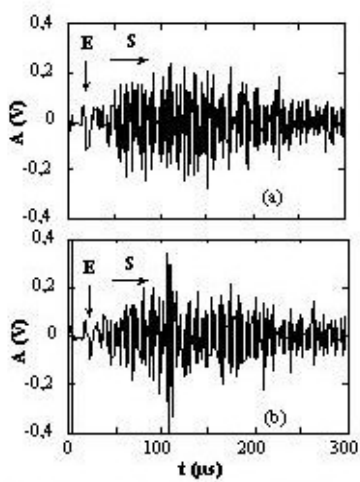

*Figure 2.3: Reproductibilité du résultat: Réponse du détecteur au même signal ultrasonore émis dans le même empilement soumis à la même contrainte* P=0.75MPa, *l'un avant, l'autre après un cycle de déchargement-rechargement: les deux signaux sont différents ce qui montre que l'empilement s'est déformé.*
*Ces signaux ont été obtenus avec un détecteur deux fois plus petit que celui de la* Fig.2.1. *Leur amplitude moyenne est plus grande que dans le cas de la Fig. 2.1. [2.4]*

Enfin, la Fig. 2.3 démontre que le signal diffusé par le milieu est différent avant et après un cycle de compression. Ceci montre que l'expérience est très sensible à la distribution des contacts et des forces de contacts ainsi qu'à leur évolution. On peut même penser utiliser ce signal pour caractériser l'évolution des contacts; en effet, cette méthode a déjà été utilisé avec les speckles optiques [2.5] et peut être adaptée aux ondes ultrasonores. On peut aussi envisager utiliser des fonctions de corrélation à plusieurs ondes pour accéder à la distribution des mouvements de grains plus complexes. On peut bâtir ainsi une spectroscopie basée sur la diffusion des ondes acoustiques.

Deux autres effets auraient pu être étudiés à l'aide de cette expérience, mais ne l'ont pas encore été: c'est tout d'abord le pic de rétro-diffusion cohérente que l'on doit observer dans la direction d'observation exactement inverse à celle de l'excitation [2.7]; pour cela il faut que l'excitation et la détection aient lieu à l'extérieur de l'échantillon.

Le second effet est la distribution des intensités diffusées dans les ondes P et les ondes S. En effet à chaque processus de diffusion, les ondes S (et P) diffusent partiellement en onde S et en onde P, dans une proportion qui dépend de la nature du diffuseur. Ainsi, au bout d'un certain nombre de processus de diffusion un équilibre statistique doit s'établir qui fixe le rapport des intensités entre ondes P et ondes S. C'est un phénomène bien connu des géo-mécaniciens des séismes.

L'équilibre statistique se calcule aisément à partir des trois considérations suivantes: (i) l'équilibre est réalisé lorsque tous les modes pouvant être peuplés sont effectivement *équi-peuplés*; (ii) le processus de diffusion conserve la fréquence; ceci implique que le spectre de fréquence se conserve après diffusion; s'il est réparti entre $\nu$ et $\nu+\delta\nu$, seuls les modes propres de fréquence entre $\nu$ et $\nu+\delta\nu$ peuvent être peuplés; (iii) le nombre de modes différents se calcule à partir d'une base quelconque; la base des fonctions $\exp[i(\mathbf{k}.\mathbf{r}-2\pi\nu t)]$ fera donc l'affaire, avec $\mathbf{k}$ (ou $\mathbf{r}$) vecteur d'onde (ou position), $|\mathbf{k}|=2\pi/\lambda=2\pi\nu/c_j$ où $c_j$ est la vitesse de propagation des ondes j=S ou P.

Pour dénombrer les modes on peut considérer un volume $V=L^3$ de matériau, chaque mode propre S ou P de vecteur d'onde $\mathbf{k}$ doit vérifier des conditions aux limites qui imposent que le produit scalaire $\mathbf{k}.\mathbf{L}$ soit égal à $2\pi(q_x+q_y+q_z)$, où les $q_m$ sont des nombres entiers quelconques. Ainsi dans la limite $\mathbf{k}.\mathbf{L}>>2\pi$ et $\delta|\mathbf{k}|=(2\pi/c_j)\delta\nu>>2\pi/L$, le nombre $\delta\rho_j$ de ces modes propres dont la fréquence est comprise entre $\nu$ et $\nu+\delta\nu$ est donné par $A_j\, q^2\, \delta q$ tels que $q=L\nu/c_j$ et $\delta q= L\, \delta\nu/c_j$, où $A_j$ est la dégénérescence de polarisation des ondes P (S); ainsi $A_j=1$ (2). On trouve donc [2.8] que le rapport de peuplement des modes est $N_S/N_P= E_S/E_P =\delta\rho_S/\delta\rho_P=2(c_p/c_s)^3$. On sait par ailleurs que le rapport $c_p/c_s$ des vitesses du son des ondes P et S est lié au coefficient de Poisson $\eta$ du matériau par $c_p/c_s=[(2-2\eta)/(1-2\eta)]^{1/2}$, *cf.* [2.9].

Mais ce rapport $N_S/N_P$ n'est pas toujours accessible directement: lorsque l'expérience consiste à détecter les ondes S et P par l'intermédiaire d'un capteur de surface S, l'expérience mesure le flux de vibration capté par la sonde; ce flux est proportionnel à $Sc_jE_j$; le rapport des intensités de vibration captées par la sonde est donc $I_s/I_p= 2(c_p/c_s)^2$, *cf.* [2.9].

## *2.2. Conclusion:*

En conclusion, la mécanique de ce système granulaire est gérée par trois longueurs différentes qu'il faut comparer: la taille de l'échantillon L, le libre parcours moyen $l_c$ et la longueur d'onde $\lambda$ de l'onde sonore. Dans les cas précédents, nous avons toujours considéré que $\lambda \ll l_c$ et $\lambda \ll L$; dans ce cas le formalisme de diffusion multiple est bien adapté, même s'il faut corriger la vitesse du son par un terme auto-cohérent [2.5]. Il en va différemment si $\lambda$ peut être nettement plus grand que $l_c$. Dans ce cas en effet, il existe des problèmes de cohérence entre les ondes diffusées par les différents diffuseurs localisés dans le même voisinage, *i.e.* $\|\mathbf{r_i}-\mathbf{r_j}\|<\lambda$. Dans le cas général, ce problème d'interférence multiple peut conduire à interdire la propagation des ondes et à provoquer le phénomène que l'on appelle localisation des ondes ou transition d'Anderson [2.4, 2.5].

Il semble cependant que ce phénomène de localisation ne puisse pas arriver pour les ondes acoustiques dans la limite des basses fréquences et des grandes longueurs d'onde; c'est une conséquence de l'existence d'un domaine de réponse quasi élastique homogène. En effet lorsque L et $\lambda$ sont grands, les déplacements le sont aussi même si la déformation est très petite. Tout point du milieu est donc obligé de suivre le mouvement de ces voisins à très basse fréquence, pour peu qu'il existe une liaison entre lui et ces derniers (matériau cohérent). S'il ne le faisait pas, c'est que cette région serait découplée mécaniquement du reste du milieu (rigidité locale infiniment faible). Ainsi, la propagation des modes de très grandes longueurs d'onde doit donc toujours être permise.

## 3. Mouvements convectifs induits dans un massif granulaire soumis à des sollicitations cycliques lentes

De nombreuses études [3.1-3.4] ont récemment eu pour but de comprendre la dynamique des milieux granulaires secs présentant une surface libre et soumis à des vibrations verticales ou horizontales. Ces milieux sont en effet le siège de multiples phénomènes lorsqu'ils sont excités dans ces conditions: lorsque la vibration est verticale par exemple, on observe la mise en tas du milieu, la formation spontanée de pentes, la génération de convection interne, la formation de structures périodiques appelées oscillons,… [3.1- 3.4]. Les vibrations horizontales engendrent, elles aussi, des mouvements convectifs [3.5].

On pourrait penser que tous ces phénomènes sont liés au caractère dynamique de l'excitation et qu'ils ne peuvent pas avoir lieu en régime quasi-statique; c'est vrai pour certains d'entre eux, (oscillons, …), mais nous voulons démontrer ici que l'apparition de mouvements de convection ne nécessite pas une excitation dynamique. Ceci peut paraître troublant a priori, car cela veut dire que la mécanique des sols quasi-statique sous conditions cycliques doit permettre la génération de courants. Or, si nous nous reportons à l'approche classique de la mécanique des sols sous comportement cyclique, le test que l'on utilise en général est le test triaxial cyclique, qui a pour but de définir l'évolution de la loi contrainte-déformation sous différentes conditions opératoires. Pour cela on se contente en général d'observer l'évolution des *conditions* *aux limites* sans chercher à savoir si l'on engendre de très grandes déformations internes. C'est pourtant ce dernier phénomène que nous avons mis en évidence il y a quelques années [3.6] concernant un talus bi-dimensionnel de rouleaux soumis à un effet bouteur cyclique (Fig. 3.1.a) dans lequel nous avons observé des mouvements extrêmement lents, convectifs (Fig. 3.1.b) et diffusifs, des grains. Nous avons ensuite reproduit cette expérience sur un massif 3-d de sable et des résultats qualitatifs analogues ont été obtenus, à savoir la mise en convection du sable; cependant le mouvement global des grains de sable est impossible à observé dans un massif 3d; c'est pourquoi nous reportons l'expérience 2d.

## *3.1. Résultats expérimentaux:*

L'expérience (Fig. 3.1.a) consiste à soumettre un massif 2-d de rouleaux (hauteur du massif H=27cm, longueur L=32,7cm, diamètre et longueur des rouleaux: d=h=5mm) à des sollicitations horizontales cycliques basses fréquences (0,3Hz-2Hz) d'amplitude A; A varie de 1 à 5cm. Quelques cylindres sont marqués de manière à pouvoir suivre leur trajectoire et à mesurer leur rotation; la très grande majorité des cylindres ont leur axe perpendiculaire au plan du massif, mais quelques autres sont parallèles à ce plan, ce qui permet de briser l'ordre triangulaire naturel de l'empilement.

Après quelques dizaines de cycles, les conditions aux limites prennent un comportement cyclique; la forme est alors celle d'un trapèze rectangle déformable (*cf.* Fig. 3.1). La poutre supérieure penche vers le bouteur, son déplacement vertical est plus important à cet endroit et plus faible près de la paroi fixe. De plus, le massif présente une surface libre dans sa partie supérieure près du bouteur, et la poutre de confinement ne repose donc pas entièrement sur lui.

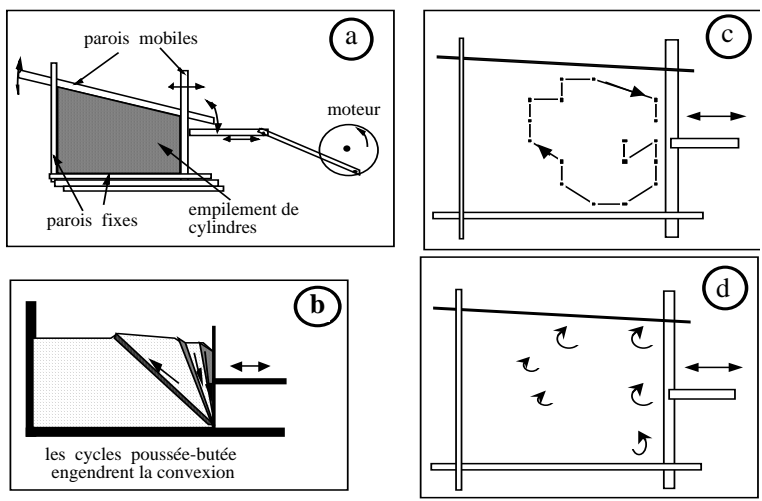

*Figure 3.1.*
*1.a: montage expérimental.*
*1.b: combinaison cyclique de poussée et de butée engendrant la convection. Dans un schéma à la Coulomb, la localisation des déformations provoque des coupures semblables à des dislocations, introduisant par là même un terme de mélange et un autre de convection.*
*1.c: Trajectoires d'un grain.*
*1.d: observation expérimentale des vitesses moyennes de rotation des grains en fonction de leur position dans le massif.*

### *convection-diffusion:*

Le mouvement alternatif du bouteur engendre un mouvement de convection et de diffusion des grains; nous donnons la trajectoire typique d'un grain en Fig. 3.1.c. Ce mouvement convectif est d'autant plus rapide que l'amplitude du déplacement du bouteur est grande (cf. paragraphe suivant). Dans les conditions expérimentales précitées, A=1cm, un grain fait un tour complet en 100 cycles à peu près. La Fig. 3.1.b récapitule l'explication que nous avions donné de ce phénomène [3.4, 3.5]; elle est basée sur le fait que le processus de déformation du massif est différent à la poussée et à la butée, ce qui génère entre autre des bandes de localisation de position et d'orientation différentes dans ces deux cas. Enfin, la diffusion observée s'explique par une certaine fluctuation de la position et de l'orientation de ces bandes de localisation, ce qui introduit un terme de coupure aléatoire et donc de mélange.

### *influence des conditions aux limites:*
Le rapport d'aspect L/H joue un rôle important à plus d'un titre: a) plus il est petit, plus le brassage est efficace. b) Il agit aussi sur la forme trapézoïdale, puisque l'inclinaison de la poutre est 13° quand L/H=1,2 et 5° quand L/H=1,87; (cette dernière valeur est sensiblement égale à celle trouvée en [3.5] pour un massif 5 fois plus petit (*i.e.* 4°)). Enfin, nous avons aussi constaté que:
i) plus l'amplitude du mouvement du bouteur est importante, plus la convection est rapide, mais cette vitesse fluctue trop au cours des cycles pour qu'on puisse en déterminer la dépendance en fonction de A; ceci nécessiterait des statistiques nombreuses.
ii) les rouleaux arrivent à diffuser dans tout le tas si l'on attend suffisamment longtemps, même pour un rapport L/H=2; cette diffusion est bien entendu d'autant plus lente que la zone considérée est éloignée du bouteur.

### *rotation des grains :*
Les grains tournent pendant leur mouvement diffuso-convectif. Nous reportons dans la Fig. 3.1.d, les vitesses moyennes observées de rotation des grains en fonction de l'emplacement dans le massif; ainsi, 6 zones typiques peuvent être définies montrant que la rotation est distribuée de façon inhomogène :
i) lorsque les grains remontent le long du plan de rupture à 45°, la rotation moyenne <R> des grains est de l'ordre de <R> ≈ $\pi/2$.
ii) cette rotation est <R> ≈ $2\pi$ pour le parcours le long de la poutre supérieure, mais cette moyenne varie fortement avec la distance à la poutre.
iii) R varie de $-\pi$ à $+8\pi$ lorsque le grain descend le long du bouteur; cet endroit est donc le lieu de très grandes fluctuations.
iv) les grains tournent en sens inverse quand ils commencent à remonter le long du plan de rupture, à la base du bouteur, <R>≈$-\alpha$, avec $\alpha$=30°±30° .
v) R≈ 0 dans le coin bas, dans le cône mort, loin du bouteur .
vi) <R> >0 près du centre de la convection; (en fait la rotation moyenne est probablement égale à celle du milieu continu équivalent, dans cette zone centrale) .
Pour fixer un ordre de grandeur, l'observation d'une dizaine de grains sur quelques cycles permet d'estimer la valeur moyenne $R_C$ de la rotation par tour de convection à $R_C$≈ $2\pi$ ±$2\pi$ .

Mais ces résultats sur les valeurs moyennes cachent de grandes disparités: en plus d'être inhomogènes, ces rotations sont intermittentes et leur sens fluctue. De plus, on constate que la bande de localisation n'est pas un lieu privilégié pour les rotations. En fait, les grains effectuent l'essentiel de leur mouvement rotatoire lorsqu'ils ne sont pas serrés les uns contre les autres et qu'ils peuvent donc rouler librement.

Ce résultat est confirmée par l'étude du mouvement des grains oblongs, dont l'axe est parallèle au plan du massif, pour lesquels la rotation est plus lente, gênée par leur plus grand encombrement stérique. C'est pourquoi aussi, les grains tournent plus rapidement près de certains bords, là où la contrainte est faible (zones proches de la poutre supérieure, près du bouteur); c'est pourquoi aussi la rotation est plus lente au sein du massif.

### 3.2 discussion et conclusion:

*Rotation et mécanique de Cosserat* [3.7]: Dans les conditions expérimentales, les mouvements des grains correspondent à des déformations "parfaitement" plastiques; il en est de même pour les rotations des grains; pour cette raison, l'énergie stockée, liée à la rotation des grains est nulle; c'est pourquoi l'application de la mécanique de Cosserat ou des matériaux micro-polaires ne semble pas du tout adaptée à notre cas expérimental. De plus, nous avons montré que le comportement rotatoire de deux grains voisins peuvent être très dissemblables ce qui se traduit par des fluctuations spatiales et temporelles importantes sur les rotations. Or ces fluctuations gouvernent la dissipation par frottement et nous n'avons aucun moyen de les prendre en compte actuellement avec le formalisme du milieu continu équivalent; ainsi, au stade où nous en sommes, la seule vraie grandeur micropolaire digne d'intérêt est la valeur moyenne de la rotation, qui est beaucoup plus faible que la somme des valeurs absolues des rotations. Enfin, cette rotation moyenne vaut approximativement $2\pi$ sur un cycle de convection complet ce qui nous suggère qu'elle est liée simplement à la rotation du référentiel lié aux lignes de courant au cours d'un cycle de convection.

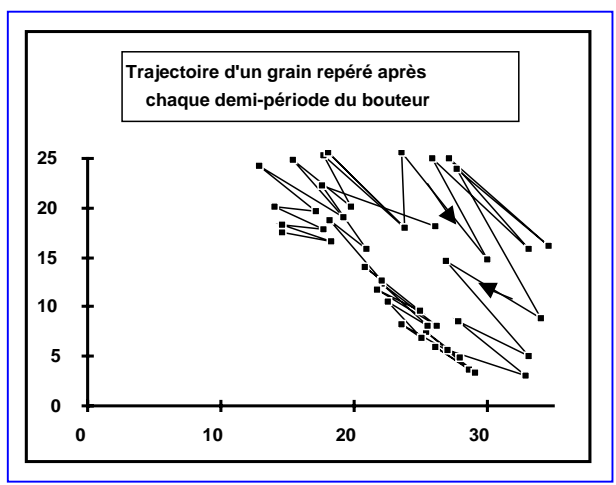

*Figure 3.2:* *Position d'un grain relevée après chaque demi période du cycle du bouteur. C'est la différence de trajectoire à la poussée et à la butée qui provoque la convection. Cette différence devient très grande près de la surface libre et près de la paroi verticale du bouteur .*

### Le mécanisme de convection:

Comme l'indique la Fig. 3.2, la convection est provoquée par la différence de mouvement des particules pendant la poussée et pendant la butée. On sait que lorsque le massif a une surface libre horizontale, l'angle de rupture est différent pendant ces deux demi-périodes (*cf.* Fig. 3.1.b); il s'ensuit un mouvement de convection et de déformation du tas. On aboutit alors à une forme stationnaire pour laquelle la surface libre du tas est inclinée. A la fin de cette évolution l'angle moyen de rupture à la poussée et à la butée doivent être identiques, c'est bien ce que l'on observe approximativement sur la Fig. 3.2. Cependant, cette mécanique est non linéaire dès que les déformations sont grandes et que le milieu change de forme. En d'autres termes, si l'on décompose les mouvements de poussée (p) et de butée (b) en 2 (ou plusieurs) étapes successives 1p+2p(+…) et (…+)2b+1b , la non linéarité de la réponse va imposer que la combinaison des mouvements va avoir une résultante non nulle et va provoquer la convection.

La Fig. 3.2 permet de mettre en évidence le rôle jouée par la paroi du bouteur et par la surface libre: c'est là que la différence des mouvements est la plus grande. En particulier, on constate que les grains en contact avec le bouteur descendent nettement après un cycle et que les angles de leur trajectoire correspondent approximativement à ceux des localisations de la Fig. 3.1.b. Ce mouvement est probablement responsable de toute la convection.

### *Parallèle avec la mécanique des fluides: cas des écoulements induits par des ondes acoustique (acoustic streaming)*

Le paragraphe précédent nous a permis de mettre en évidence que l'évolution des conditions aux limites au cours d'un cycle était responsable de la convection. Il nous semble important de faire un parallèle entre ce processus de convection et ceux que l'on rencontre lorsqu'un fluide contenu dans un récipient est soumis à des ondes sonores ou ultrasonores [3.8]. Dans ce dernier cas, on sait que le mouvement vibratoire imposée aux parois induit dans le liquide des mouvements oscillants. Le mouvement liquide peut être considéré comme non visqueux loin de la paroi, mais près de celle-ci ce n'est plus le cas. L'épaisseur $\delta e$ de cette couche visqueuse dépend de la fréquence $\omega/(2\pi)$ de l'onde et de la viscosité cinématique $\nu$ du fluide, suivant la loi $\delta e = (2\nu/\omega)^{1/2}$. A cause de cette viscosité, et aux grandes amplitudes de vibration, le mouvement du fluide au voisinage de la paroi présente un déphasage par rapport au mouvement des parois, ce qui engendre une convection.

De même ici, la Fig. 3.1 montre que le mouvement des grains proches de la paroi ne suivent pas le mouvement de la paroi, mais descendent systématiquement.

Il est intéressant de noter qu'un régime de convection presque similaire à celui de la Fig. 3.1 est observé pour des milieux granulaires contenus dans un récipient soumis à des vibrations horizontales à plus haute fréquence [3.5]. Dans ce cas, le régime d'excitation n'est plus quasi-statique, car l'on observe un décollement périodique de la paroi vertical du container. Mais ce mouvement périodique de décollement-recollement joue un rôle similaire au mouvement du bouteur dans notre cas. Ainsi, on est amené à faire le lien entre les régimes de convection dans les milieux granulaires soumis à des mouvements périodiques en quasi-statique et en dynamique et entre ceux-ci et les mouvements de convection induits par des vibrations dans les fluides.

Une différence subsiste probablement entre les deux régimes quasi-statique et dynamique: dans le premier cas le mouvement semble confiné dans une épaisseur égale approximativement à la hauteur de la paroi verticale, ce qui ne semble pas vérifié dans le second cas; cette différence est probablement liée aux forces mises en jeu: dans le premier cas la force appliquée est localisée près de la paroi, tandis que dans le second cas, l'accélération périodique est une force qui peut agir sur l'ensemble du volume et peut le rendre plastique en tout point.

*Parallèle avec la mécanique des fluides: l'advection chaotique*

L'expérience de la Fig. 3.1 génère de grandes fluctuations. Mais, la nature même des fluctuations observées et leur impact sur les propriétés mécaniques du matériau restent posés: ces fluctuations sont-elles réellement aléatoires ou au contraire chaotiques avec des intermittences, des distributions spatiales et temporelles très larges et d'amplitudes variées? Pour mieux cerné ce problème, il est intéressant de poursuivre le parallèle entre ce problème et celui de la convection dans un liquide bi-dimensionnel.

En effet, on peut faire un parallèle, au sens strict, entre ce problème de convection dans les milieux granulaires et le problème d'un écoulement stationnaire dans un fluide quelconque [3.9]. Supposons en effet que ce mouvement, une fois moyenné sur plusieurs périodes, se résume à un mouvement de convection; il se caractérisera par des lignes de courant stationnaire, se fermant sur elles-mêmes. Plaçons nous en représentation Lagrangienne et appelons **v** la vitesse locale du fluide; le problème étant bidimensionnel et le milieu étant quasi incompressible, on peut écrire div(**v**)=0; ceci impose que l'on peut décrire à partir d'une fonction $\psi$ telle que:

$$dx/dt = \partial \psi / \partial y \quad \& \quad dy/dt = - \partial \psi / \partial x \qquad (3.1)$$

où $\psi$ est appelé « fonction courant ».

Formellement, le problème devient équivalent au problème d'une particule dans un champ, qui est gérée par le système d'équation:

$$dp/dt = \partial \mathbf{H} / \partial q \quad \& \quad dq/dt = - \partial \mathbf{H} / \partial p \qquad (3.2)$$

où **H** est la fonction de Hamilton de la particule. On sait que ce problème est intégrable et qu'il ne peut donner lieu à du chaos. Ainsi, un écoulement bidimensionnel permanent d'un fluide incompressible ne peut donner lieu à du chaos et ne peut donc pas donner lieu à du mélange. Pour qu'il y ait mélange, il faut rajouter des termes de diffusion, qui sont générés à l'échelle locale par des échanges de particules de temps typique $\tau_o$. Le temps caractéristique $\tau_{diff}$ de la diffusion sur un échantillon de taille L avec des particules de taille d (ou des sauts de longueur d) varie comme $\tau_{diff} \sim \tau_o (L/d)^2$.

Ceci impose donc que le temps de mélange doit croître énormément lorsqu'on accroît la taille du système. L'analogie ne serait plus valable si l'écoulement était tri-dimensionnel; un tel écoulement pourrait alors donner lieu à du chaos et à un terme de mélange beaucoup plus efficace.

*Fluctuations du mouvement des grains:*

Notre expérience a mis en évidence l'existence de fluctuations importantes des mouvements, qui s'ajoutent au terme convectif. Le problème est de savoir si ces fluctuations perturbent fondamentalement la convection ou si elles rajoutent un terme diffusif simple. Pour bien comprendre la nature de ce problème et ses enjeux, il est intéressant de le reformuler de la manière suivante:

Dans le paragraphe précédent, nous avons montré qu'un milieu bidimensionnel soumis à un mouvement convectif permanent mélangeait très mal, car la dynamique du mélange était caractérisée par un coefficient de diffusion microscopique. Ceci conduit à estimer que le temps caractéristique $\tau_L$ de mélange doit croître comme le carré du rapport entre la taille de l'expérience (L) et la taille du grain d: $\tau_L \sim \tau_o (L/d)^2$. Par contre, nous avons émis l'hypothèse que les fluctuations que nous voyons dans cette expérience sont, en partie au moins, liées à la fluctuation de la position et de la direction de la bande de localisation. Cette part des fluctuations provient des fluctuations des paramètres macroscopiques (frottement solide, densité initiale, frottement des parois,…) [3.5, 3.10]. Dans ces conditions, cette partie des fluctuations peut être fonction de la taille L de l'expérience et de l'amplitude A des oscillations, et non de la taille d des grains; dans ce cas, le coefficient de diffusion par cycle sera proportionnel à la taille L de l'expérience, et non à la taille d du grain, et le temps $\tau_L$ de mélange et de diffusion sera indépendant de la taille de l'expérience. L'autre partie des fluctuations, celle qui reste reliée à des processus réellement microscopiques, qui font intervenir le mouvement relatif de deux grains voisins, cette autre partie devra induire un mécanisme de mélange dont la rapidité décroît avec la taille du système comme $1/\tau_L \sim (d/L)^2/\tau_o$.

A ce stade, on doit rappeler que la diffusion des molécules dans un liquide est le mécanisme de mouvement le plus intense à l'échelle local, mais qu'il reste totalement inopérant à l'échelle macroscopique (par rapport à la convection).

Ainsi, l'analyse des fluctuations, et de leur régression en fonction de la taille du système, est donc essentielle pour comprendre la mécanique sous-jacente. Elle est de plus fondamentale pour asseoir les bases d'une physique statistique propre aux milieux granulaires. C'est ce dernier point que nous voulons montrer dans le paragraphe suivant.

*Intérêt de l'expérience:*
Cette expérience montre que les sollicitations cycliques imposées à un milieu granulaire permettent de changer la configuration des grains les uns par rapport aux autres, de changer la configuration des contacts et la configuration des forces locales. Ceci doit permettre le traitement statistique de ces distributions et de retenir comme état réel l'une des configurations les plus probables. C'est donc la base du traitement statistique de la mécanique des milieux granulaires.

Nous avons vu cependant que le cas bidimensionnel est un cas pathologique, car la convection n'y est pas chaotique; dans ces conditions, c'est la diffusion qui permet le transfert des particules d'une ligne de courant à l'autre et qui gère le mélange à grande distance. Cette diffusion est-elle normale ou anormale, les données expérimentales ne permettent pas encore de conclure. Cependant, cette discussion laisse entrevoir qu'un milieu granulaire peut présenter une perte importante de la mémoire de ses conditions initiales, ce qui permet de justifier ainsi l'application du théorème ergodique, sous certaines conditions comme d'attendre suffisamment longtemps.

Ce qui est valable pour un milieu bidimensionnel doit l'être, bien entendu, dans le cas tridimensionnel, qui est encore plus favorable. Dans ces conditions il est probable qu'une nouvelle compréhension de la mécanique des milieux granulaires passe par la compréhension des termes convectifs et diffusifs que nous venons de mettre en évidence. Un autre champ d'application de cette approche est le problème de la ségrégation des particules dans un milieu granulaire. Nous en donnerons un exemple plus tard.. Mais avant d'aborder ce dernier problème, il est important de parfaire l'analogie qu'il y a entre les problèmes de mécanique des fluides et de mécanique des milieux granulaires en démontrant qu'un milieu granulaire saturé de liquide et soumis à des vibrations très intenses peut se comporter strictement comme un liquide non visqueux. La raison à cela est que les forces inertielles deviennent prépondérantes dans des conditions de vibration intense et la loi rhéologique devient négligeable.

## 4. Comportement "fluide parfait" d'un milieu granulaire saturé de liquide et soumis à des vibrations intenses

Dans cette section, nous allons montrer qu'un milieu granulaire peut présenter une surface libre qui se déforme de façon équivalente à celle d'un liquide parfait sous certaines conditions. Pour cela il suffit d'imposer des vibrations telles que les forces inertielles qu'elles engendrent soient nettement plus importantes que les forces de pesanteur. Dans ces conditions, nous allons montré qu'on peut engendrer un phénomène similaire à la houle à la surface du sable; un point remarquable est que la surface du sable s'aplanit dès qu'on arrête l'excitation, comme s'aplanit la mer lorsque le vent retombe. Comme dans le cas de la houle, le mécanisme qui engendre la déformation de la surface du sable est l'instabilité de Kelvin-Helmholtz qui force l'interface entre deux fluides en mouvement relatif à ne pas rester plane mais à onduler; cependant, dans l'expérience que nous allons décrire, le relief qui est engendré ne se propage pas, contrairement à ce qui se passe pour la houle.

   Nous commencerons par présenter le phénomène, puis nous donnerons les bases de mécanique des fluides nécessaires pour sa compréhension. Enfin nous montrerons comment ces mécanismes hydrodynamiques peuvent être combinés et utilisés soit pour contrôler la vitesse de sédimentation soit pour gérer la position des interfaces entre phases en apesanteur. Il est probable que ces deux domaines ne sont encore qu'au début de leur caractérisation et qu'ils ont un potentiel d'applications importants .

### *4.1. Une expérience simple:*

Prenons une cellule fermée rectangulaire (longueur L, hauteur h ≤ L, épaisseur e<L) ou cylindrique de rayon R=D/2, d'axe horizontal. Remplissons la à moitié de sable puis saturons la complètement d'alcool ou d'eau. Plaçons cette cellule sur un vibreur qui vibre horizontalement (amplitude b, fréquence f=$\omega/(2\pi)$; b sin$\omega$t); (axe de vibration parallèle à la longueur). Lorsque la fréquence est suffisante (>5Hz), on observe que l'interface entre le sable et l'eau devient ondulée lorsque l'amplitude et la fréquence de vibration dépasse une valeur seuil. Cette ondulation est stationnaire dans le référentiel de la cellule; c'est à dire qu'elle ne dépend pas du temps et qu'elle ne se propage pas. C'est un relief figé dans le référentiel de la cellule; nous le caractériserons par son amplitude a et sa longueur d'onde $\lambda=2\pi/k$, ou son nombre d'onde k. Par la suite, on exprimera ces grandeurs en unités réduites A=a/h (ou a/D) et K=hk (ou Dk).

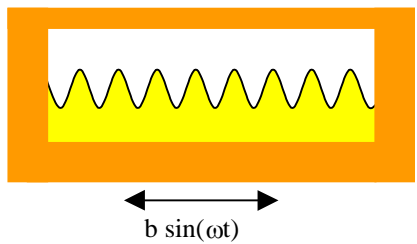

*Figure 4.1: Relief apparaissant à l'interface liquide-sable dans une cellule vibrée horizontalement. L'interface ne dépend pas du temps: c'est un relief ; mais il s'efface dès qu'on arrête la vibration.*

Les caractéristiques a et λ du relief dépendent de b et de ω , comme le montre la Fig. 4.2.a. Un autre point important est que le relief disparaît, et l'interface redevient plane et horizontale, dès qu'on repasse au dessous du seuil d'apparition du relief, ou que l'on coupe la vibration. Les reliefs de la Fig. 4.2.a ont été obtenus dans une cellule cylindrique et pour le couple sable-alcool éthylique; on voit aussi que a et λ sont toujours proportionnels. De plus, la Fig. 4.2.b qui reporte la variation de K=2πD/λ en fonction de W=(b²ω²)/gD pour différentes valeurs de b et de ω démontre que seul le produit bω intervient dans l'amplitude A et la longueur d'onde λ. De plus on observe que A varie proportionnellement à λ tant que A est suffisamment petit devant la hauteur de la cellule h . Nous verrons par la suite que le paramètre important est le carré de la vitesse typique W=b²ω².

En fait, si l'on regarde plus précisément le relief, on s'aperçoit qu'il est animé d'un mouvement oscillant horizontal rapide, *i.e.* à la fréquence égale à celle de l'excitation; l'amplitude $a_h$ de ce mouvement horizontal est très faible $a_h \ll \lambda$.

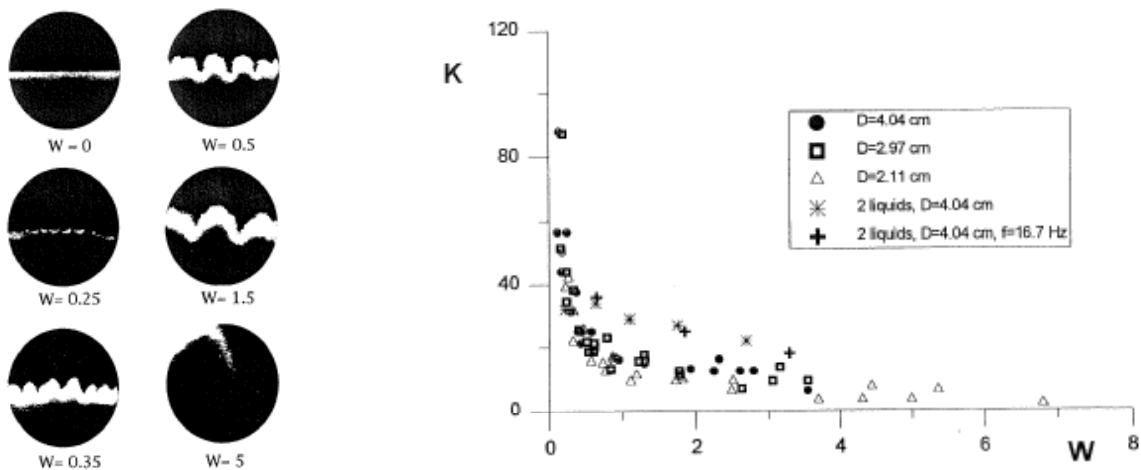

*Figure 4.2: a: reliefs typiques obtenus pour certaines valeur de W=b²ω²/(gD). D=diamètre de la cellule.*
*b: Dépendance du nombre d'onde K=2πD/λ du relief en fonction de W=b²ω²/(gD). Les valeurs correspondent au couple sable-alcool éthylique. En fait , on trouve que K~1/W . Ainsi, λ varie linéairement avec W. On a ajouté dans b une courbe qui provient de reliefs formés entre 2 liquides (ici fluorinert et huile de ricin).*

Nous allons maintenant chercher à comprendre le phénomène est à décrire le ou les mécanismes mis en jeu. Pour cela nous devons faire quelques rappels d'hydrodynamique.

## *4.2. Quelques rappels d'hydrodynamique:*

### *épaisseur de la couche limite visqueuse:*

Lorsqu'on fait osciller à une fréquence $\omega/(2\pi)$ une plaque plane parallèlement à sa surface, et que cette plaque est immergée dans un liquide visqueux (densité $\rho$, viscosité dynamique $\eta$, viscosité cinématique $\nu=\eta/\rho$), le liquide est entraîné par le mouvement de la plaque sur une épaisseur $\delta=(2\nu/\omega)^{1/2}$. $\delta$ est appelée épaisseur de la couche limite visqueuse. Ainsi, l'amplitude d'oscillation du fluide décroît de façon exponentielle avec la distance r à la plaque. L'entraînement du fluide par la plaque est une conséquence de l'existence des forces visqueuses. $\delta$ est d'autant plus petite que la fréquence $\omega/(2\pi)$ est grande. On peut donc considérer que le liquide n'est pas entraîné par le mouvement de la plaque au delà de l'épaisseur $\delta$ ; loin de la plaque, le liquide se comporte donc comme s'il était parfait, c'est-à-dire non visqueux.

### *Théorème de Bernoulli:*

Si l'on considère l'écoulement stationnaire d'un fluide parfait (non visqueux), il se caractérise par des lignes de courant. Un tel écoulement obéit au théorème de Bernoulli; celui-ci stipule que la somme de l'énergie cinétique et de la pression est constante le long d'une ligne de courant. Soit

$$\rho v^2/2 + p = c^{ste} \tag{4.1}$$

Ce théorème est important, car il explique beaucoup de phénomènes simples, comme le fait qu'une feuille de papier reposant sur une table et sur laquelle on souffle parallèlement à la surface de la table est maintenue plaquée sur cette table et ne s'envole pas; il permet aussi d'expliquer l'instabilité de Kelvin-Helmholtz, comme nous allons le voir.

### *Instabilité de Kelvin-Helmholtz:*

Considérons deux liquides immiscibles de densité différente $\rho_1$ et $\rho_2$ en contact le long de leur surface plane commune; le plus lourd ($\rho_2>\rho_1$) est en dessous du plus léger; supposons qu'ils bougent l'un par rapport à l'autre, mais qu'ils restent en contact; pour des raisons de continuité, leur vitesse perpendiculairement à l'interface doit être la même de part et d'autre de l'interface, mais leurs vitesses tangentielles (parallèlement à l'interface) sont différentes.

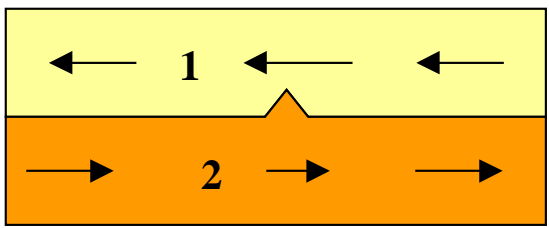

*Figure 4.3:* *l'interface plane de 2 liquides se déplaçant à des vitesses différentes est instable : toute déformation verticale vers le haut ou vers le bas provoque une variation de pression qui l'amplifie.*

Dans ces conditions, cherchons à savoir si l'interface reste plane ou non; pour cela, plaçons-nous dans le référentiel barycentrique et considérons une déformation de la surface comme celle représentée sur la Fig. 4.3. Si l'interface s'est déformée localement vers le haut; la conservation du débit impose que la vitesse du fluide 1, $v_1$, augmente au niveau de l'excroissance et que celle, $v_2$, du fluide 2 diminue; d'après le théorème de Bernoulli, ceci impose que la pression locale $p_1$ décroît au dessus de la bosse et que $p_2$ augmente. Il s'ensuit que la bosse est aspirée vers le haut; elle a donc tendance à s'amplifier spontanément. Un raisonnement analogue montre qu'un trou crée une surpression au dessous de lui et qu'il est aussi amplifié; de même, ce phénomène est indépendant du sens et de la direction du mouvement. Une analyse de stabilité linéaire montre que le processus est instable quelque soit la longueur d'onde, et qu'il est d'autant plus rapide que la longueur d'onde est petite (p. 154 de la ref. [4.2]).

En fait un calcul exacte dans le cas de deux liquides doit tenir compte des forces capillaires. Kelvin a trouvé dans ce cas, *cf.* p. 351 problème 3 de [4.2], qu'il existe une différence de vitesse $v_1-v_2=U$ minimum entre les deux liquides $U_{KH}$ au dessous de laquelle la surface plane reste stable; mais celle-ci devient instable lorsque U atteint ou dépasse $U_{KH}$ ; il apparaît alors un mode plus instable, caractérisée par une longueur d'onde précise $\lambda_{KH}$. $U_{KH}$ et $\lambda_{KH}$ sont données par :

$$\|v_1-v_2\|^2 = U^2_{KH} = 2[(\rho_1+\rho_2)/(\rho_1\rho_2)] \, [\alpha g(\rho_2-\rho_1)]^{1/2} \qquad (4.2.a)$$

$$\lambda_{KH} = 2\pi \, [\alpha/\{g(\rho_2-\rho_1)\}]^{1/2} \qquad (4.2.b)$$

où $\alpha$ est la tension superficielle entre les deux liquides. L'Eq. (4.2.a) confirme que l'instabilité apparaît dès que la différence de vitesse est non nulle pour des liquides sans tension superficielle ($\alpha=0$), et l'Eq. (4.2b) que la longueur d'onde la plus instable est $\lambda=0$ dans ce cas.

Cette instabilité est appelée l'instabilité de Kelvin-Helmoltz ; c'est elle qui produit la houle sur la mer dès que le vent est suffisamment intense.

Ces équations ne décrivent pas la forme du relief pour autant : pour cela il faudrait faire un calcul non linéaire recherchant la forme d'équilibre. D'un point de vue physique, on comprend que la hauteur $a(\lambda)$ du relief doit saturer pour un $\lambda$ donné à une valeur proche de $\lambda$, *i.e.* $a \approx \lambda$.

### *4.3. cas du mouvement oscillant; interprétation des résultats sur le sable*

Pour comprendre ce qui se passe dans le cas d'un mouvement oscillant, nous allons partir des Eq. (4.2) correspondant à un cas en présence de forces capillaires et suivre l'approche proposée en [4.3]. Nous appliquerons ensuite les résultats au cas où les forces capillaires sont nulles.

Il résulte des forces d'inertie et de la différence de densité des deux liquides que, lorsque la cellule est animée d'un mouvement périodique $b\sin\omega t$, celle-ci communique aux deux liquides un mouvement oscillant caractérisé par leur différence de vitesse U :

$$U = b\omega \cos\omega t \, (\rho-1)/(\rho+\chi) \qquad (4.3)$$

où $\rho=\rho_2/\rho_1$ est le rapport des densités et où $\chi = h_2/h_1$ est le rapport des hauteurs des deux couches liquides, avec $h=h_1+h_2$.

Comme nous avons vu que le mécanisme de l'instabilité de Kelvin-Helmholtz ne dépendait pas du sens du courant, le relief poursuivra son développement de demi-période en demi-période. Le seuil d'instabilité correspondra donc probablement aux conditions 4.2 pour lesquelles on aura choisi le carré de la différence de vitesse U² égale à la moyenne du carré de la différence de vitesse donnée par l'Eq. (4.3). Dans ces conditions on trouve :

$$(b^2\omega^2)_{\text{seuil KH}} = 4\,[\alpha g/(\rho_2-\rho_1)]^{1/2}\,[(\rho+1)(\rho+\chi)^2]/[\rho(\rho-1)(1+\chi)^2] \qquad (4.4.a)$$

La découverte expérimentale de ce phénomène est probablement due à Wolf [4.4] ; Lyubimov et Cherepanov [4.5] ont effectué un calcul plus complet que celui qui est exposé ici, basé sur un développement à plusieurs échelles de temps, l'un rapide correspondant à la fréquence d'excitation, les autres plus lents, qui intègrent le mouvement sur plusieurs périodes. Le calcul de [4.5] correspond au cas $\chi=1$; il confirme la valeur du seuil précédent, mais l'étend aussi au cas $\alpha=0$, absence de tension superficielle. Ces auteurs trouvent:

$$(b\omega)^2 = \{(\rho_1+\rho_2)^3/[2\rho_1\rho_2\,(\rho_2-\rho_1)]\}\,\{\alpha k+(\rho_2-\rho_1)g/k\}\,\text{th}(kh) \qquad (4.4.b)$$

avec $k=2\pi/\lambda$.

Ce calcul montre aussi que la bifurcation qui correspond au passage du relief plan au relief sinusoïdal est critique, au sens des bifurcations, c'est-à-dire que l'amplitude $a_r$ du relief varie près du seuil comme :

$$a_r \propto [\,(b^2\omega^2)-(b^2\omega^2)_{\text{seuil KH}}\,]^{1/2} \qquad (4.5.a)$$

$$\text{où } A=a_r/h \propto [\,(b^2\omega^2)-(b^2\omega^2)_{\text{seuil KH}}\,]^{1/2}/(gh) \qquad (4.5.b)$$

Dans le cas de deux liquides ($\alpha\neq 0$), nous avons montré dans [4.3] que la hauteur du relief varie linéairement avec $(b^2\omega^2)-(b^2\omega^2)_{\text{seuil}}$, à la précision de nos mesures; elle ne suit donc pas l'Eq. (4.5). Nous supposons que c'est parce que nous n'étions pas suffisamment près du seuil; en effet dans les cas étudiés, le rapport $a_r/\lambda$ était de l'ordre de 1 ; or l'Eq. (4.5) pour être valable suppose que a est très petit, c'est-à-dire $a_r/\lambda\ll 1$. Nous avons aussi observé que $\lambda$ reste constant près du seuil. L'accroissement de $\lambda$ passe par un doublement de période.

Enfin, beaucoup plus loin du seuil, on observe que les variations de $\lambda$ suivent celles de l'amplitude $a_r$. L'explication de ces trois derniers phénomènes est la suivante: lorsque l'on est proche du seuil, la longueur capillaire $l_{\text{capillaire}}=\{\alpha/[g(\rho_2-\rho_1)]\}^{1/2}$ joue un rôle important. C'est elle qui fixe la longueur d'onde $\lambda$ du relief: $\lambda=2\pi\,l_{\text{capillaire}}$. Au fur et à mesure que le relief croît, l'amplitude $a_r$ devient vite plus grande que $l_{\text{capillaire}}$ ; la longueur d'onde initiale est déstabilisée; on observe expérimentalement un mécanisme de doublement de période [4.3].

Quand la hauteur $a_r$ du relief devient grande devant $l_{\text{capillaire}}$, $\lambda$ le devient aussi. Il s'ensuit que $l_{\text{capillaire}}$ ne reste plus la longueur physique de référence. Dans ces conditions, on peut négliger $\alpha$ dans l'Eq. (4.4.b). Si $\lambda$ reste petit devant h, $\text{th}(kh)\approx 1$ , l'Eq. (4.4.b) conduit à :

$$\lambda = 2\pi\,\{(b\omega)^2/g\}\,[2\rho_1\rho_2(\rho_2-\rho_1)]/(\rho_1+\rho_2)^3 \qquad (4.5.c)$$

$a_r$ et $\lambda$ croissent à peu près de la même façon. On peut se servir de h comme unité de mesure; dans ce cas l'Eq. (4.5.c) devient:

$$\lambda/h = 2\pi \ \{[2\rho_1\rho_2(\rho_2-\rho_1)] / (\rho_1+\rho_2)^3\} \ \{(b\omega)^2/(gh)\} \qquad (4.5.d)$$

Ceci correspond aux comportements de la Fig. 4.2.

Lorsque le relief atteint une hauteur comparable à h et que la longueur d'onde devient plus grande que h, th(kh) devient petit: th(kh)∝1/(kh). Ceci n'arrive qu'aux très fortes amplitudes et α reste donc totalement négligeable, *i.e.* α=0. Dans ces conditions la résolution de l'Eq. (4.4.b) conduit au relief:

$$\lambda/h = 4\pi \ \{b\omega/(gh)^{1/2}\} \ [\rho_1\rho_2(\rho_2-\rho_1)]^{1/2} / (\rho_1+\rho_2)^{3/2} \qquad (4.5.e)$$

Lorsque la hauteur du relief devient plus grande que la taille de la cellule, le relief se détruit et les deux liquides se séparent ; leur surface de séparation devient alors plane et verticale. Cette nouvelle configuration d'équilibre peut se comprendre aussi relativement simplement, si l'on remarque que dans cette position les deux fluides n'ont plus besoin, ni de raison de bouger l'un par rapport à l'autre: ils sont en équilibre mécanique.

Pour résumer, l'Eq. (4.5) résume les différents cas de figure possible. Les Eqs. (4.5.a) & (4.5.b) s'appliquent très près du seuil; les Eqs (4.5.c) & (4.5.d) nettement au dessus du seuil, ou en l'absence de tension superficielle. Enfin l'Eq. (4.5.e) s'applique lorsque la hauteur du relief atteint la hauteur du container.

### *4.4. Robustesse du mécanisme de l'instabilité:*

L'apparition d'un relief gelé, sinusoïdal ou en dents de scie, semble un mécanisme d'instabilité très robuste, puisque nous l'avons même observé dans le cas d'un équilibre de $CO_2$ diphasique gaz-liquide près de son point critique [4.6], *cf.* Fig. (4.4). Or dans ce cas, le fluide est hypercompressible.

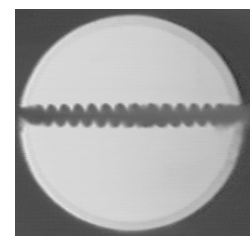

*Figure 4.4 : $CO_2$ biphasique sous vibration horizontale près de $T_c$ (c'est-à-dire hypercompressible) : génération d'un relief permanent stable qui dépend des paramètres de la vibration (a,ω) et de (T-$T_c$). Le phénomène est identique à celui que l'on observe pour les biphasiques liquide-liquide ou liquide-sable.*

### *Cas du mélange liquide-sable :*

Il est bien évident que la tension superficielle à l'interface liquide-sable est extrêmement faible, voire nulle, puisque le sable est déjà tout entier mouillé par le liquide. Dans ces conditions l'instabilité devrait apparaître dès les premières vibrations et suivre la dépendance décrite par les Eqs. (4.5.d) & (4.5.e).

En fait, il existe plusieurs limitations qui empêchent de voir ce phénomène aux vitesses de vibration très faibles:

- Tout d'abord, le sable doit être liquéfié ; cela nécessite un cisaillement cyclique suffisamment important, et donc une vibration horizontale suffisamment intense, correspondant à des accélérations $b\omega^2$ de l'ordre de ou supérieures à g.
- Ensuite, la longueur d'onde λ du relief doit être bien supérieure au diamètre d du grain.
- Mais cela ne suffit pas encore, car il faut que les grains agissent de conserve et que le milieu granulaire ressemble à un milieu incompressible, qui ne se dilate pas. Ce dernier point n'est rendu possible que grâce aux interactions hydrodynamiques et ceci requiert que la viscosité du liquide soit suffisamment grande, de telle sorte que l'épaisseur δ de la couche limite visqueuse soit au moins de l'ordre de la taille d'un pore. Soit en première approximation :

$$\delta=(2\nu/\omega)^{1/2} > d/7 \qquad (4.6)$$

Cette condition semble particulièrement importante, car nous n'avons pu détecter l'instabilité de Kelvin-Helmoltz pour un mélange $CO_2$–sable. Sans en être certain, nous attribuons ce phénomène à la viscosité trop faible du $CO_2$, qui devient incapable de maintenir les grains en une phase cohérente.

### 4.5. Conclusion : autres applications potentielles

Enfin, nous voudrions terminer cette section en montrant le potentiel d'application des vibrations, du théorème de Bernoulli et de l'instabilité de Kelvin-Helmholtz à des cas spécifiques tels que le contrôle des interfaces en apesanteur. En effet, en apesanteur la position des interfaces entre fluides est imprévisible; la gestion des fluides multiphasés y pose donc un défi technique important. Or nous avons vu que les vibrations , grâce à l'instabilité de Kelvin-Helmholtz, doit permettre d'orienter les interfaces perpendiculairement à la direction de vibration dès que le paramètre de vibration $W=b^2\omega^2$ est grand devant gh. Ce résultat reste valable en apesanteur; et le contrôle des vibrations devient ainsi un enjeu capital. Cependant, il existe toujours des micro-fluctuations de gravité effective dans les vaisseaux spatiaux, (station orbitale, navette spatiale, fusée sonde); ces micro-fluctuations sont produites soit par le mouvement des cosmonautes, soit par les autres expériences, soit encore par des perturbations extérieures (atmosphère résiduelle, vent solaire,…); ainsi, tant que l'on ne contrôlera pas la direction préférentielle de vibration dans l'espace, le comportement des interfaces semblera défier la logique. En contrepartie, imposer des vibrations devrait permettre de contrôler ces interfaces. C'est ce que nous voulons montrer par les deux expériences suivantes:

### *Vibration d'un mélange liquide-gaz de $CO_2$ près de $T_c$ en apesanteur:*

Lors d'une expérience de vibration en fusée sonde (Mini Texus 5, février 1998) sur du $CO_2$ diphasique près de son point critique, nous avons observé une structuration de l'interface perpendiculaires à la direction de vibration comme le montre la Fig. 4.5; or la densité des deux phases sont pourtant très voisines. Ceci démontre l'importance capitale de ces phénomènes pour l'hydrodynamique spatiale.

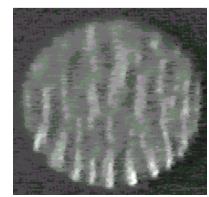

*Figure 4.5 : $CO_2$ diphasique près de $T_c$, en apesanteur, et soumis à des vibrations « horizontales » ; on observe que le milieu se stratifie perpendiculairement à la direction de vibration*

## Structuration en couche d'un lit fluidisé qui sédimente sous vibration verticale:

D'autres exemples, aussi dignes d'intérêt, sont décrits dans la ref. [4.7]. On y explique en particulier comment orienter en strates monocouches horizontales séparées les unes des autres un milieu granulaire qui sédimente : pour cela, il suffit d'utiliser des vibrations verticales. On constate aussi dans ce cas que la vitesse de sédimentation y est très fortement diminuée ; c'est probablement en partie lié à des effets de parois…

L'application du théorème de Bernoulli permet de donner une explication simple de ces phénomènes. C'est la diminution de pression imposée par Bernoulli qui attire une particule vers les parois orientées parallèlement à la direction de vibration ou qui aligne les particules perpendiculairement à la direction de vibration, *cf.* [4.7].

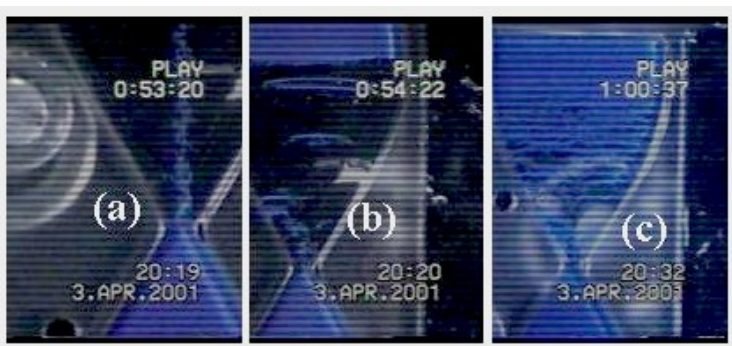

*Figure 4.6 : Stratification d'une sédimentation sous l'action de vibrations verticales.*
*Cas de billes moins denses que le liquide*
*(a) : sédimentation sans vibration ;*
*(b) et (c) : sédimentation sous vibration verticale intense.*

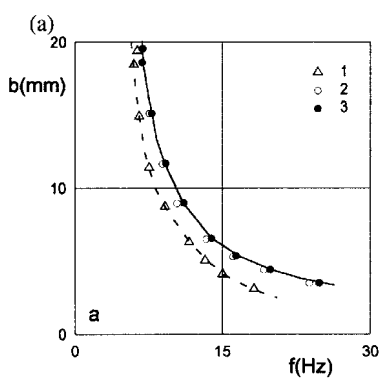 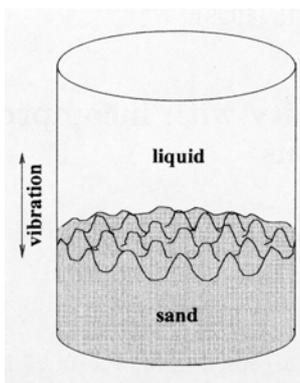 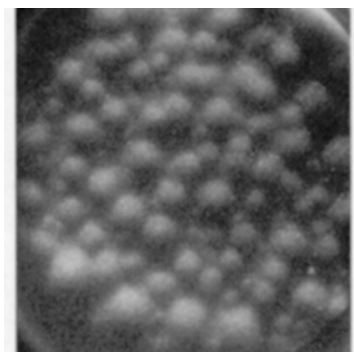

*Figure 4.7: **Effet des vibration verticales** b sin(2πft) sur un mélange liquide-billes de verre (diamètre d=0.1mm): (au milieu) vue générale ; (à droite) : relief en forme de collines ; ce régime est obtenu entre les courbes (1) et (3) de la Fig. de gauche . (à gauche) les seuils de transition : En dessous de la courbe (1) la surface plane reste stable ; elle se déforme au dessus de (1) est il apparaît un relief permanent en forme de collines ; la taille moyenne des collines croît avec l'amplitude de vibration, jusqu'à ce qu'on atteigne la courbe (2) ; au dessus de (2) des crises d'oscillation paramétriques apparaissent, qui détruisent le haut des collines par intermittences ; au dessus de (3) la liquéfaction de la surface est complète et son oscillation paramétrique permanente.*

Pour être complet, il faudrait aussi développer ce qui se passe lorsque l'on vibre verticalement un milieu granulaire saturé de liquide sur terre [4.8]. On peut y engendrer un relief de collines, puis à plus forte intensité la liquéfaction de la surface du sable, qui est alors le siège d' ondes paramétriques… Nous donnons la limite entre les régimes dans la Figure 4.7, ainsi qu'un exemple de collines ; ces collines restent visibles et non déformées après l'arrêt de la vibration.

## 5. Mélange et ségrégation dans un milieu granulaire

Un des problèmes majeurs de l'industrie des poudres et des milieux granulaires est celui de leur mélange et de leur ségrégation [5.1-5.6]. C'est pourquoi de nombreux articles traitent depuis longtemps de ce sujet : par exemple des mélangeurs utilisant des cylindres tournant autour de leur axe horizontal ont été très étudiés; la ségrégation axiale qui y est engendrée après quelques heures est connue au moins depuis 1939 [5.7]. Plus récemment, certains physiciens ont redécouvert ce problème, *cf.* articles récents de *Science*, *Nature* ; ils y ont même vu une gageure fondamentale de la physique [5.8]. Cela a permis de développer des études par Imagerie par Résonance Magnétique (IRM) des ségrégations radiale et axiale en cylindres tournants [5.9] ou l'étude des écoulements granulaires [5.10-5.11].

Le but de cette section n'est pas de décrire toutes les facettes de la ségrégation, mais d'essayer d'ancrer ce problème à celui de la chimie-physique via la théorie des mélanges [3.9], problème que nous avons déjà aborder dans la section 3. En effet, très peu de scientifiques, à l'exception de quelques théoriciens ou expérimentateurs du génie chimique n'ont encore fait ce parallèle [5.12 ; 5.13]; pourtant ces derniers développent cette approche depuis fort longtemps.

Comme nous l'avons vu dans la section 4.3, on peut faire le parallèle entre le problème d'un écoulement bi-dimensionnel permanent à volume constant et un problème hamiltonien décrivant le mouvement à une dimension d'une particule. Il en résulte qu'un  système mécanique engendrant un écoulement bi-dimensionnel permanent à volume constant ne peut être un bon mélangeur, car le seul processus de mélange qui peut y être induit est celui d'une diffusion ; or on sait que celle-ci n'est efficace qu'à courte portée. Pour s'en convaincre, on remarquera (i) que les lignes de courants d'un régime stationnaire 2d forment des boucles concentriques de rayon la taille du récipient, et (ii) que seule la diffusion permet le mélange perpendiculairement aux lignes de courants.

Autrement dit, pour être efficace, un mélangeur doit nécessairement imposer un écoulement non permanent et induire du chaos et de la turbulence, sinon il mélange mal; et donc rien d'étonnant à ce qu'il ségrège. Un exemple typique de ces systèmes est le cylindre tournant. Pour pallier ce problème, nous avons opté d'étudier la ségrégation dans un système considéré comme turbulent et chaotique, appelé le Turbula® (Willy A. Bachofen AG Maschinenfabrik, Basel, Switzerland). Cet appareil est couramment utilisé dans les laboratoires de développements pharmaceutiques ; il engendre un mouvement complexe de la cellule, qui est la combinaison de deux mouvements de rotations et d'une translation ; ceci entraîne un flux complexe des particules à l'intérieur du récipient, voir licence [5.14] pour une description détaillée du mouvement. Le détail des expériences que nous reportons peuvent être trouvées dans [5.15].

Les billes que nous avons utilisées sont des billes de sucre (NP Pharm, France) de diamètres différents, variant de 0,35 mm à 1,1 mm, mais de même densité. Certaines billes de 1mm ($d_{ref}$=1 mm), appelées par la suite "billes de références", ont été dopées avec une huile organique pour être visible par IRM ; ce dopage n'a pas changé leur propriétés d'écoulement (frottement solide, compacité, dilatance,….). Le rapport des diamètres R= $d_{ref}$/d a été varié de R=0,9 à R=2,8. Ces billes sont disposées en proportions adéquates dans un récipient cylindrique (diamètre 52 mm, hauteur 66mm) que l'on peut placer tour à tour dans le Turbula® et dans un spectromètre IRM (type Bruker DSX100 (100MHz)) de manière à permettre d'étudier l'évolution de la distribution des billes marquées au fur et à mesure du nombre de tour $N_R$ imposé par le Turbula®.

## *5.1. Résultats expérimentaux*

On commence par remplir le récipient ($N_R$=0) au 2/3 en deux couches horizontales de grains; les billes de référence sont en bas, les autres en haut. L'échantillon est donc dans un état initial complètement ségrégé. Une première image IRM est alors acquise ; puis on refait une image à chaque tour de Turbula® . Ces images 3d sont en fait constituées de séries de 16 images 2-d: si Oz est l'axe du récipient, deux séries de 16 images sont obtenues, l'une correspond à une série de coupes horizontales parallèles au plan xy (coupes axiales), l'autre à des coupes verticales parallèles au plan zy (coupes sagittales). L'épaisseur d'une coupe étant de 3,5 mm, les 16 tranches correspondent à une épaisseur totale de 56 mm. Chaque coupe 2d contient 64*64 pixels de résolution spatiale de 940 µm. L'intensité du voxel est proportionnelle au nombre de billes de référence (dopées) qu'il contient. Par conséquent, la distribution spatiale de l'intensité des voxels permet de déterminer celle de la concentration en grains dopés. Un rapport signal/bruit adapté nécessite un temps d'acquisition de l'ordre de 20 minutes pour l'ensemble des images.

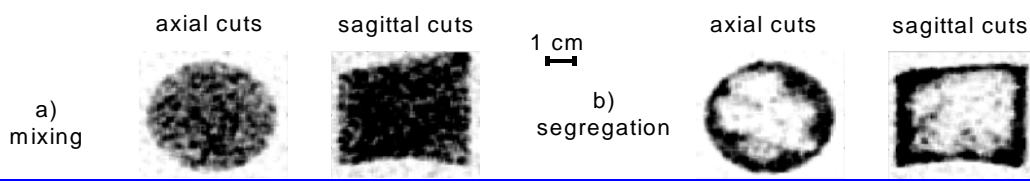

*Figure 5.1:* *Distribution stationnaire obtenue après* 220 *rotations à* 22 tr/mm. *La cellule a été remplie au 2/3 : a) Mélange de billes identiques* (R=1). *b) Ségrégation de billes de taille différentes* (R=2.8)*; la région des plus grosses billes apparaît foncée.*

Après un nombre suffisant de révolutions (220 tours), une analyse directe des images IRM permet de distinguer deux états stationnaires différents selon que R=1 ou R≠1 (Figs. 5.1.a & 5.1.b). Pour les billes similaires, un mélange quasi parfait est observé, ce qui confirme en passant que les propriétés rhéologiques des billes dopées et non dopées sont les mêmes. Mais une ségrégation apparaît pour des billes de tailles différentes, les plus grosses étant localisées près de la paroi interne du récipient.

Une analyse quantitative nécessite la définition d'un indice de ségrégation S ; il en existe de nombreuses définitions. Celle que nous avons choisie est basé sur l'approche statistique de Lacey [5.16] ; c'est le rapport $S = \sigma/\bar{x}$ entre l'écart-type et la moyenne de la distribution de l'intensité des voxels. Avec cette définition $S(N_R)$ varie de 1 (pour une ségrégation totale) à $1/\sqrt{b}$ (pour un mélange parfaitement homogène) pour un mélange binaire (50-50%), où b est le nombre moyen de grains par voxel. L'analyse quantitative des expériences de mélange/ségrégation est reportée sur la Fig. 5.2.a qui montre les variations de S en fonction du nombre de tours $N_R$. Pour le mélange (R=1), S décroît de 0,95 à 0,48 quand le nombre de révolutions $N_R$ croît de 0 à 30. Par ailleurs, S varie aléatoirement entre 0,92 et 0,95 au lieu de 1 quand $N_R=0$ ; cette incertitude sur S, qui est de l'ordre de 10% est liée aux effets de surface et d'interface entre les deux couches de billes.

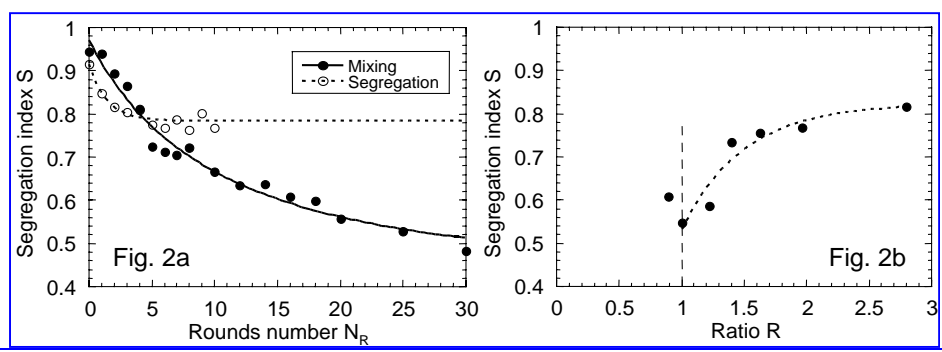

***Figure 5.2 :*** *Indice de ségrégation S obtenu à une vitesse de rotation de $\Omega=22$ tr/mn ; les points sont les valeurs expérimentales: a) variation de S vs. $N_R$, pour le mélange de particules identiques et la ségrégation. Les courbes représentent les modélisations par des décroissances mono-exponentielles. b) variations de S en fonction de R obtenu après un grand nombre de tours $N_R$ ($N_R=220$) ; les valeurs de S pour R=0.9 et pour R=1.2 sont semblables. On remarque que la ségrégation est déjà totalement développée dès R=1.4.*

Ainsi, la valeur stationnaire S=0,48 pour le mélange est obtenue dès $N_R=30$. Pour la ségrégation (R=2,8), S évolue de 0,92 à 0,78 ; cette dernière valeur est atteinte dès $N_R=5$ . Les temps caractéristiques des cinétiques de mélange et de ségrégation ont été obtenus en ajustant par une loi exponentielle les points expérimentaux de la Fig. 5.2.a. Les temps caractéristiques sont respectivement 1,4 et 10,7 révolutions pour la ségrégation (R=2,8) et le mélange (R=1). Ainsi, la ségrégation est un phénomène beaucoup plus rapide que le mélange dans un Turbula®. C'est pourquoi la ségrégation s'observe facilement avec ce montage. On a reporté dans la Fig. (5.2.b) les variations de S à l'état stationnaire (*i.e.* $N_R=220$ cycles) en fonction de $R=d_{ref}/d$. En l'absence d'effets de taille finie, on s'attend à $S(R) \approx S(1/R)$, c'est ce qui est observé ici, au moins dans la limite de la précision expérimentale, pour les deux points R=0,9 et R=1,2. De plus, les échantillons R>1 et R<1 présentent des motifs de ségrégation similaires puisque les plus grosses billes sont toujours localisées sur les bords alors que les plus petites sont concentrées au centre du récipient. Enfin, on constate que la ségrégation est pleinement développée dès R=1,4. Ces expériences démontrent donc l'efficacité du Turbula® comme dispositif à séparer des particules relativement très peu différentes, ce qui n'était pas prévu a priori.

En ce qui concerne la ségrégation entre particule de taille très proche, R≈1, *i.e.* R=0.9 ou 1.1, un résultat très semblable a été obtenu par simulation numérique dans un cylindre tournant [5.17]. Ainsi, bien que le turbula® soit un meilleur mélangeur que le cylindre tournant, cela ne suffit pas à limiter l'efficacité de la ségrégation. Ceci est lié à la rapidité du mécanisme de ségrégation ( $\tau$=1.4 tr).

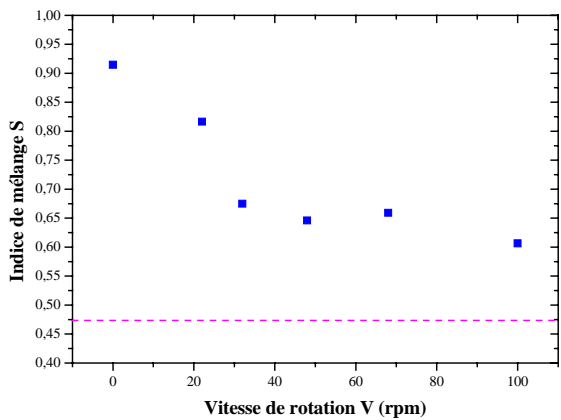

*Figure 5.3 : Variation de l'indice de ségrégation en fonction de la vitesse $\Omega$ (ou V) de rotation du turbula.*

*La ségrégation diminue quand l'amplitude a de vibration imposée par le mouvement de translation du turbula devient plus grande que la gravité g, i.e. $L_1\Omega^2>g$ ; $L_1$ est de l'ordre de la distance séparant les axes du turbula.*

## *5.2. Interprétation*

Un des mécanismes principaux de ségrégation est la percolation des petits grains à travers un lit de gros grains cisaillés. Cette percolation est induite par les forces de pesanteur qui permettent aux petits grains de tomber « verticalement » dans les lacunes laissées par les gros. Ce mécanisme apparaît dès que le milieu granulaire est déformé par cisaillement ou en écoulement . C'est pourquoi il est très efficace dans le cas où le turbula , où il a lieu à la surface libre. Ce mécanisme exclut les petits grains du pourtour extérieur du milieu granulaire ; il est efficace quelque soit la vitesse $\Omega$ de rotation du turbula.

Nous allons voir qu'un autre mécanisme force ces même grains à ce concentrer au centre lorsque la vitesse du turbula est faible. C'est probablement ce mécanisme qui engendre la variation de l'indice de ségrégation en fonction de $\Omega$.

## *5.3. Mouvements de particules isolées*

On peut étudier le mouvement d'un grain de taille différente dans une mer d'autres grains en fonction de la différence de taille entre le grain isolé et les autres grains. Pour que l'étude expérimentale soit facilitée, il faut utiliser comme grain sonde une graine, (graine de pavot par exemple), qui donne un fort signal RMN et qui est donc facilement détectable. Le résultat d'une telle étude est reportée sur la Fig. 5.4, lorsque la vitesse du turbula® est relativement rapide ($\Omega \approx$60 tr/mn). On y voit en particulier que les grains plus petits que la taille de la mer se déplacent au hasard dans cette mer, sans toutefois aborder la surface. Au contraire, un gros grain se retrouve très vite à la périphérie du milieu.

Si l'on diminue la vitesse de rotation $\Omega$, *i.e.* $\Omega$ =22tr/mn par exemple, on constate que la dynamique d'un gros grain isolé reste la même, et qu'il reste localisé à l'extérieur (D=1) ; par contre, on constate que la dynamique d'un petit grain converge vers le centre de l'échantillon, où il reste confiné au bout d'un certain nombre de rotations. Ainsi, il existe un second mécanisme de ségrégation, qui ne s'applique qu'aux petits grains dans une mer de gros grains, et à basse vitesse $\Omega$.

C'est pourquoi l'on peut conclure que le turbula® mélange correctement une petite quantité de petits grains dans une mer de gros grains, pour peu que la vitesse de rotation du turbula® soit grande, et qu'on puisse négliger les effets de bords, ce qui est le cas si le rapport surface/volume est petit. Ce résultat est valable pour des concentrations pouvant aller jusqu'à moins 20%. La proportion inverse, une petite quantité de gros grains dans une mer de petits grains, n'est pas bien mélangée.

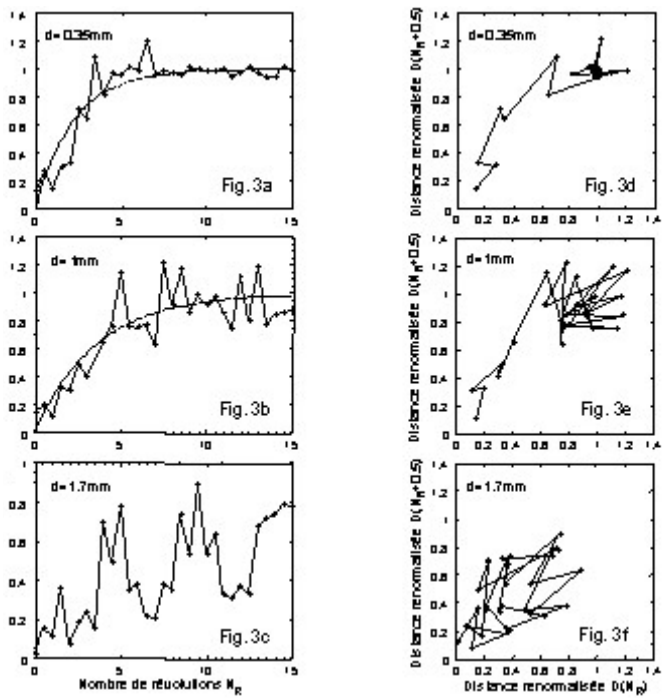

*Figure 5.4 : Evolution de la position d'une graine de pavot dans une mer de billes de sucres, dans le turbula.*
*(a-c) Évolution de la distance renormalisée au centre de l'échantillon $D(N_R)$ de la graine en fonction du nombre de tours $N_R$ du turbula, à $\Omega$=22 tr/min , pour des billes de sucre de : (a) 0.35mm, (b) 1 mm, et (c) 1.7 mm.*
*Les courbes des Figures (a) et (b) sont des ajustements mono-exponentiels.*
*(d-f) Évolution de $D(N_R+0.5)$ versus $D(N_R)$ pour un milieu constitué de billes de sucre de (d) 0,35 mm, (e) 1 mm, et (f) 1.7 mm.*

Ce second mécanisme de ségrégation diminue lorsque la vitesse de rotation augmente, car l'augmentation de vitesse engendre des forces centrifuges et des effets inertiels supplémentaires qui vibrent le milieu dans plusieurs directions à la fois ; ceci perturbe violemment l'écoulement et augmente l'efficacité du processus de mélange. Ce processus « de brassage » est contrôlé par au moins deux nombres sans dimensions relié aux caractéristiques expérimentales $M_1= L_1\Omega^2/g$ et $M_1= L_2\Omega^2/g$, où $L_1$ et $L_2$ sont la taille du container et la distance séparant les axes de rotation du turbula respectivement. C'est pourquoi l'indice de ségrégation diminue lorsqu'on augmente la vitesse du turbula . C'est ce qui est effectivement observé sur la Fig. 5.3.

Lorsqu'on ajoute une faible quantité de fluide au mélange hétéro-granulaire, le problème change totalement de nature, car des forces de cohésion apparaissent qui permettent aux grains de s'agglomérer, et le turbula® redevient un excellent mélangeur. Des études ont montré cependant que l'homogénéisation du milieu pouvait demander un temps très conséquent.

## 6. Conclusion

Nous avons cherché à montrer à travers quelques exemples, que le comportement dynamique des matériaux granulaires pouvait être très diverse, allant d'une réponse analogue à celui d'un solide (propagation du son) à une réponse analogue à celui d'un liquide parfait (instabilité de Kelvin-Helmholtz), c'est-à-dire non visqueux. Il peut même se comporter comme un biphasique constitués de deux ou plusieurs phases granulaires "immiscibles" (ségrégation), et présenter l'analogue d'une transition de phase "amas"-"gaz granulaire".

Tous ces problèmes peuvent être traités d'un point de vue très pratique, car ils ont des incidences dans de nombreux procédés industriels. Mais ils soulèvent aussi un grand nombre de questions fondamentales, qui permettent ou permettront d'éclairer d'un jour nouveau un certain nombre d'effets et d'approches.

Les milieux granulaires sont donc le siège d'un certain nombre de comportements non linéaires, qui peuvent rendre leur comportement global surprenant et complexe. A notre avis, cette complexité du comportement spatio-temporel global n'est pas la preuve de l'absence de règles simples du comportement macroscopique local; mais seulement la preuve que ces règles sont non linéaires. C'est en tout cas ce que laisse croire les quelques exemples proposés dans cet article: en effet nous avons pu interpréter le comportement du matériau dans chacun des cas présentés en ne faisant intervenir que des notions et des comportements simples.

Bien entendu les exemples proposés ne sont sûrement qu'un petit aperçu des différentes facettes de la dynamique des milieux granulaires et ils n'ont aucun caractère exhaustif. Un point qui aurait mérité d'être développé est celui de l'importance des conditions initiales, et donc de la mécanique quasi statique, sur le développement des instabilités tels que les avalanches, les ruptures,…

## Références

The electronic arXiv.org version of this paper has been settled during a stay at the Kavli Institute of Theoretical Physics of the University of California at Santa Barbara (KITP-UCSB), in june 2005, supported in part by the National Science Fundation under Grant n° PHY99-07949.

*Poudres & Grains* can be found at :
http://www.mssmat.ecp.fr/rubrique.php3?id_rubrique=402